\documentclass[12pt]{article}
\usepackage{amssymb}
\usepackage{graphics}
\usepackage{epsfig}
\usepackage{a4wide}

\begin{document}
\newcommand{\ppzh}{$pp \to Z^0H^0 +X~$ }
\newcommand{\qqzh}{$q\bar q \to Z^0H^0~$}
\newcommand{\qqzhg}{$q\bar q \to Z^0H^0g~$}
\newcommand{\qgzhq}{$q(\bar q)g \to Z^0H^0q(\bar q)~$}
\newcommand{\ppuuzh}{$pp \to u\bar u \to Z^0H^0+X~$}
\newcommand{\ppqqzh}{$pp \to q\bar q \to Z^0H^0+X~$}

\title{ Associated $Z^0H^0$ production with leptonic decays at LHC in next-to-leading order QCD }
\author{ Zhang Shi-Ming, Zhang Ren-You, Ma Wen-Gan, and Guo Lei \\
{\small Department of Modern Physics, University of Science and Technology}\\
{\small of China (USTC), Hefei, Anhui 230026, P.R.China}  }

\date{}
\maketitle \vskip 15mm

\begin{abstract}
In this work we investigate the effects of the littlest Higgs model
(LHM) up to the QCD next-to-leading order (NLO) on the $Z^0H^0$
associated production at the CERN Large Hadron Collider (LHC). We
study the dependences of the leading order and NLO QCD
corrected integrated cross sections for this process on the
factorization/renormalization scale and the LHM parameters. We also
provide the distributions of the transverse momenta of final decay
products $\mu^-$ and $\tau^-$. Our results show that the heavy
neutral gauge bosons $Z_H$ and $A_H$ could induce significant
discrepancies from the standard model predictions. It is found that
when the LHM parameters are taken as $c=0.5$, $c^{\prime}=0.22$,
$f=4~TeV$ and $\mu=(M_H+M_Z)/2$, the effects at the
$\sqrt{s}=14~TeV$ LHC from the heavy neutral gauge boson are about
$12.83\%$ and $10.37\%$ to the leading order and NLO QCD corrected integrated
cross sections, respectively. We also conclude that the NLO QCD
corrections at the $\sqrt{s}=14~TeV$ LHC can obviously reduce the
scale uncertainty of the integrated cross section, and significantly
enhance the differential cross sections of $p_T^{\mu^-}$ and
$p_T^{\tau^-}$. It demonstrates that the precision measurement of
the $Z^0H^0$ associated production process at the LHC could provide
the clue of the LHM physics.
\end{abstract}


\vfill \eject

\baselineskip=0.32in

\renewcommand{\theequation}{\arabic{section}.\arabic{equation}}
\renewcommand{\thesection}{\Roman{section}.}
\newcommand{\nb}{\nonumber}

\newcommand{\Dir}{\kern -6.4pt\Big{/}}
\newcommand{\Dirin}{\kern -10.4pt\Big{/}\kern 4.4pt}
\newcommand{\DDir}{\kern -7.6pt\Big{/}}
\newcommand{\DGir}{\kern -6.0pt\Big{/}}

\makeatletter      
\@addtoreset{equation}{section}
\makeatother       

\par
\section{Introduction}
\par
The CDF and D0 experiments ruled out the standard model (SM)
\cite{sm1,sm2} Higgs boson with mass between $156~GeV$ and $177~GeV$
at $95\%$ confidence level (CL) \cite{mh-Fermilab}. Recently, the
ATLAS and CMS experiments at the CERN Large Hadron Collider (LHC)
have excluded most of the Higgs mass ranges of
 $146-466~GeV$ and
$145-400~GeV$ at $95\%$ CL in their reports of \cite{mh-Atlas} and
\cite{mh-CMS}, respectively. Currently, the ATLAS and CMS groups
exclude a substantial region of the possible Higgs boson mass range,
and find several Higgs like events around the locations of $M_{H}
\sim 126~GeV$ (ATLAS) and $M_{H} \sim 124~GeV$ (CMS)
\cite{mh-Atlas-1,mh-CMS-1}. Further searching for Higgs boson and
studying its properties are still the important tasks for the
present and upcoming high energy colliders.

\par
Despite the tremendous success of the SM in describing the high
energy physics at the energy scale up to several hundred $GeV$, the
instability of the Higgs boson mass leads to the ``hierarchy
problem`` \cite{hie} which comes from the quadratic loop corrections
to the Higgs boson mass. In order to give a proper electroweak
symmetry breaking (EWSB) scale, the Higgs boson mass needs unnatural
fine-tuning when it gets a radiative correction with the cutoff
scale about $10~TeV$. In order to solve the ``hierarchy problem``,
physicists developed several new particle models such as
supersymmetry \cite{super}, extra dimensions \cite{extra}, little
Higgs \cite{lh1,lh2}, technicolor and so on. Among these theories,
the little Higgs models are proposed as one kind of models without
fine-tuning in which the Higgs boson is naturally light as a result
of nonlinearly realized symmetry. The littlest Higgs model (LHM)
\cite{lhest1,lhest2, lhest3} is the most economical model of them
and a phenomenological viable model.

\par
There are an $SU(5)$ global symmetry and a locally gauged subgroup
$G_{1} \otimes G_{2}=[SU(2)_{1} \otimes U(1)_{1}] \otimes [SU(2)_{2}
\otimes U(1)_{2}]$ in the LHM. At the scale $\Lambda_{S}$, the
global symmetry $SU(5)$ is broken into its subgroup $SO(5)$. At the
same time, the local gauge symmetry $[SU(2) \otimes U(1)]^{2}$ is
also spontaneously broken into its diagonal subgroup
$SU(2)_{L}\otimes U(1)_{Y}$, which is identified as the SM gauge
group. In the LHM, a set of new heavy gauge bosons ($W_{H}^{\pm}$,
$Z_{H}$ and $A_{H}$) and a new heavy-vector-like quark ($T$) are
introduced to cancel the quadratic divergence induced by SM gauge
boson loops and the top quark loop, respectively. These new gauge
bosons might provide the significant signatures at the present and
future high energy colliders. The \ppzh process is one of the main
production mechanisms of Higgs boson with moderate mass at the LHC,
which gives a very distinctive signature. This process could be used
to measure the Higgs mass and the couplings between Higgs boson and
gauge bosons and determine the quantum numbers of the Higgs boson.
Therefore, investigating the process \ppzh at the LHC in the context
of the LHM is necessary for probing the LHM physics \cite{w3}. We
find that the SM and minimal supersymmetric standard model (MSSM) 
analyses to the \ppzh
process at the LHC have been already existed in
Ref.\cite{smwork}.

\par
In this work we study the effects of the LHM on neutral Higgs boson
production associated with $Z^0$ boson up to the QCD next-to-leading
order (NLO) at the CERN LHC. In the LHM the new neutral gauge
bosons, such as $Z_{H}$ and $A_{H}$, give additional contributions
to this process. The paper is constructed as follows: In section II,
we provide related theory of the LHM to our calculations. In section
III, we describe the calculations at the leading order (LO) and the 
QCD NLO for the
\ppzh process. The numerical results and discussions are presented
in section IV. Finally, a short summary is given.

\par
\section{Related theory of LHM}
\par
The LHM is based on an $SU(5)/SO(5)$ nonlinear $\sigma$ model. The
vacuum expectation value (VEV) breaks the $SU(5)$ global symmetry
into its subgroup $SO(5)$ and breaks the local gauge symmetry
$[SU(2)\otimes U(1)]^2$ into its diagonal subgroup $SU(2)_L\otimes
U(1)_Y$ at the same time, which is identified as the SM electroweak
gauge group. The gauge fields $W^{\prime \mu}$ and $B^{\prime \mu}$
associated with the broken gauge symmetries are related to the SM
gauge fields by
\begin{equation}
W^{\prime\mu}=-cW_{1}^{\mu}+sW_{2}^{\mu},\hspace{1cm}
W^{\mu}=sW_{1}^{\mu}+cW_{2}^{\mu},
\end{equation}
\begin{equation}
B^{\prime \mu}=-c^{\prime}B_{1}^{\mu}+s^{\prime}B_{2}^{\mu},
\hspace{1cm} B^{\mu}=s^{\prime}B_{1}^{\mu}+c^{\prime}B_{2}^{\mu},
\end{equation}
with mixing angles of
\begin{equation}
c=\frac{g_{1}}{\sqrt{g_{1}^{2}+g_{2}^{2}}},\hspace{2.5cm}
c^{\prime}=\frac{g_{1}^{\prime}}{\sqrt{g_{1}^{\prime2}+g_{2}^{\prime2}}}.
\end{equation}
At the scale $f$ the SM gauge bosons remain massless, while the
heavy gauge bosons acquire masses of order $f$. The $W$ and $B$ are
identified as the SM gauge bosons, with couplings of $g=g_1 s=g_2 c$
and $g^{\prime}=g_1^{\prime} s^{\prime}=g_2^{\prime} c^{\prime}$.
The electroweak symmetry breaking (EWSB) gives the masses for the SM
gauge bosons and induces further mixing between the light and heavy
gauge bosons. We denote the light gauge boson mass eigenstates as
$W^{\pm}(W_L^{\pm})$, $Z^0(Z_L^{0})$ and $\gamma(A_L)$ and the new
heavy gauge boson mass eigenstates as $W_{H}^{\pm}$, $Z_{H}$ and
$A_{H}$. The masses of the charged and neutral gauge bosons to the
order of $v^2/f^2$ are given by \cite{lhest1}
\begin{eqnarray}\label{WL-mass}
M_{W^{\pm}}^2=M_{W_L^{\pm}}^2 &=& m_w^2 \left[1 - \frac{v^2}{f^2}
\left( \frac{1}{6}+ \frac{1}{4} (c^2-s^2)^2 \right) + 4
\frac{v^{\prime 2 }}{v^2}\right],
\end{eqnarray}
\begin{eqnarray}\label{WH-mass}
M_{W_H^{\pm}}^2 &=& m_w^2\left( \frac{f^2}{s^2c^2v^2}-1\right) ,
\label{MWH}
\end{eqnarray}
\begin{eqnarray}\label{AH-mass}
M_{\gamma}^{2}=0,\hspace{0.5cm}
M_{A_{H}}^{2}=m_{z}^{2}S_{W}^{2}\left(\frac{f^{2}}{5s^{\prime2}c^{\prime2}v^{2}}-1
+\frac{\chi_{H}C_{W}^{2}}{4s^{2}c^{2}S_{W}^{2}}\right),
\end{eqnarray}
\begin{eqnarray}\label{ZL-mass}
M_{Z}^{2}=M_{Z_{L}}^{2}=m_{z}^{2}\left\{1-\frac{v^{2}}{f^{2}}\left[\frac{1}{6}
+\frac{1}{4}(c^{2}-s^{2})^{2}+
\frac{5}{4}(c^{\prime2}-s^{\prime2})^{2}-\frac{\chi^{2}}{2}\right]\right\},
\end{eqnarray}
\begin{eqnarray}\label{ZH-mass}
M_{Z_{H}}^{2}=m_{z}^{2}C_{W}^{2}\left(\frac{f^{2}}{s^{2}c^{2}v^{2}}-1
-\frac{\chi_{H}S_{W}^{2}}{s^{\prime2}c^{\prime2}C_{W}^{2}}\right),
\end{eqnarray}
with
\begin{eqnarray}
\chi=\frac{4fv^{\prime}}{v^{2}}, \hspace{1cm}
\chi_{H}=\frac{5S_{W}C_{W}}{2}\frac{scs^{\prime}c^{\prime}(c^{2}s^{\prime2}
+s^{2}c^{\prime2})}{5C_{W}^{2}s^{\prime2}c^{\prime2}-S_{W}^{2}s^{2}c^{2}},
\end{eqnarray}
where $m_z\equiv gv/(2C_W)$, $C_W\equiv
\cos\theta_W=\frac{m_w}{m_z}$, $\theta_{W}$ is the Weinberg angle,
$v^{\prime}$ and $v$ are the VEV's of the scalar $SU(2)_{L}$ triplet
and doublet, respectively. In the following numerical calculations
we take $v = 246~GeV$ and $\chi = 0.5$.

\par
The couplings of the neutral gauge bosons to quarks are expressed in
the form as $i\gamma_{\mu}(g_L P_L+g_R P_R)$ where $P_{L,R}\equiv
\frac{1}{2}(1\mp \gamma_5)$. The explicit expressions are given
below.
\begin{equation}\label{Z-coupling-1}
g_{L}^{Z\bar U U}=-\frac{e}{2S_{W}C_{W}}\left
\{1-\frac{4}{3}S_{W}^{2} +\frac{v^{2}}{f^{2}}\left[\frac{c^{2}}{2}
(c^{2}-s^{2})-\frac{5}{2}(c^{\prime2}-s^{\prime2})\left(\frac{8}{15}-\frac{1}{3}c'^{2}\right)
\right]\right\},
\end{equation}
\begin{equation}\label{Z-coupling-2}
g_{R}^{Z\bar U
U}=-\frac{e}{2S_{W}C_{W}}\left\{-\frac{4}{3}S_{W}^{2}-\frac{v^{2}}
{f^{2}}\left[\frac{5}{2}(c^{\prime2}-s^{\prime2})
\left(\frac{2}{15}+\frac{2}{3}c'^{2}\right)\right]\right\},
\end{equation}
\begin{equation}\label{Z-coupling-3}
g_{L}^{Z\bar D
D}=-\frac{e}{2S_{W}C_{W}}\left\{-1+\frac{2}{3}S_{W}^{2}-\frac{v^{2}}{f^{2}}
\left[\frac{c^{2}}{2}
\left(c^{2}-s^{2}\right)+\frac{5}{2}\left(c^{\prime2}-s^{\prime2}\right)\left(-\frac{2}{15}
+\frac{1}{3}c^{\prime2}\right)\right]\right\},
\end{equation}
\begin{equation}\label{Z-coupling-4}
g_{R}^{Z\bar D D}=-\frac{e}{2S_{W}C_{W}}\left\{\frac{2}{3}S_{W}^{2}
-\frac{v^{2}}{f^{2}}\left[\frac{5}{2}(c^{\prime2}-s^{\prime2})
\left(\frac{4}{15}-\frac{2}{3}c'^{2}\right)\right]\right\},
\end{equation}
\begin{eqnarray}\label{ZH-coupling-1}
g_{L}^{A_{H}\bar U
U}=\frac{e}{2s^{\prime}c^{\prime}C_{W}}\left(\frac{2}{15}-\frac{1}{3}c'^{2}\right),\hspace{0.5cm}
g_{R}^{A_{H}\bar U
U}=\frac{e}{2s^{\prime}c^{\prime}C_{W}}\left(\frac{8}{15}-\frac{8}{6}c^{\prime2}\right),
\end{eqnarray}
\begin{eqnarray}\label{ZH-coupling-2}
g_{L}^{A_{H}\bar D
D}=\frac{e}{2s^{\prime}c^{\prime}C_{W}}\left(\frac{2}{15}-\frac{2}{6}c^{\prime2}\right),\hspace{0.5cm}
g_{R}^{A_{H}\bar D
D}=\frac{e}{2s^{\prime}c^{\prime}C_{W}}\left(-\frac{4}{15}+\frac{4}{6}c^{\prime2}\right),
\end{eqnarray}
\begin{eqnarray}\label{ZH-coupling-3}
g_{L}^{Z_{H}\bar U U}=\frac{ec}{2s S_{W}},\hspace{0.5cm}
g_{R}^{Z_{H}\bar U U}=0,\hspace{0.5cm} g_{L}^{Z_{H}\bar D
D}=-\frac{ec}{2s S_{W}}, \hspace{0.5cm} g_{R}^{Z_{H}\bar D D}=0,
\end{eqnarray}
where $U$ and $D$ represent the up-type $(U=u,c,t)$ and
down-type $(D=d,s,b)$ quarks, respectively. The couplings between
neutral gauge boson and Higgs boson are expressed as
\begin{equation}
g^{HZZ}=\frac{ie^{2}v g_{\mu\nu}}{2S_{W}^{2}C_{W}^{2}}
\left\{1-\frac{v^{2}}{f^{2}}\left[\frac{1}{3}-\frac{3}{4}\chi^{2}+\frac{1}{2}(c^{2}-s^{2})^{2}+
\frac{5}{2}(c^{\prime2}-s^{\prime2})^{2}\right]\right\},
\end{equation}
\begin{equation}\label{Z_H-A_H-ZH}
g^{HZA_{H}}=-\frac{ie^{2}v
g_{\mu\nu}}{2S_{W}C_{W}^2}\frac{c^{\prime2}-s^{\prime2}}{2s^{\prime}c^{\prime}},
\hspace{0.8cm} g^{HZZ_{H}}=-\frac{ie^{2}v
g_{\mu\nu}}{2S_{W}^2C_{W}}\frac{c^{2}-s^{2}}{2sc}.
\end{equation}

\par
The heavy neutral gauge boson $V_{H}~(V_H=Z_H,A_H)$ can decay into a
fermion pair and $Z^0H^0$. We obtain the partial decay rates
expressed below \cite{width}.
\begin{equation} \label{VH-width-1}
\Gamma(V_{H}\to f\bar f)=\frac{N_{c}}{12\pi}
\left[(g^{V_{H}\bar f f}_{v})^{2}(1+2r_{f})+(g^{V_{H}\bar f f}_{a})^2(1-4r_{f})\right]\sqrt{1-4r_{f}}M_{V_{H}},
\end{equation}
\begin{equation}\label{VH-width-2}
\Gamma(V_{H}\to
Z^0H^0)=\frac{(g^{V_H})^2}{192\pi}\sqrt{\lambda}
\left[(1+r_{Z}-r_{H})^{2}+8r_{Z}\right]M_{V_{H}},
\end{equation}
where $N_c = 3$ is the color factor, $g^{V_Hf \bar f}_v=(g^{V_Hf \bar f}_R+g^{V_Hf \bar f}_L)/2$,
$g^{V_Hf \bar f}_a=(g^{V_Hf \bar f}_R-g^{V_Hf \bar f}_L)/2$,
$g^{A_H}=g^{\prime}(c^{\prime 2}-s^{\prime 2})/(2c^{\prime}s^{\prime})$,
$g^{Z_H}=g(c^{2}-s^{2})/(2cs)$, $\lambda=1+r_{Z}^2+r_{H}^2-2r_{Z}-2r_{H}-2r_{Z}r_{H}$,
and $r_{i}=X_{i}^2/M^2_{V_{H}}$ $(X_{i}=m_f,M_Z,M_H)$.
Since in our investigated parameter space the $V_H \to T\overline{T}$ and $V_H
\to \overline{T}t(T\bar{t})$ decays are kinematically forbidden, we
assume that the total decay width $\Gamma_{V_H}~(V_H=Z_H,A_H)$ is
the sum of $\Gamma(V_{H}\to f\bar f)$ and $\Gamma(V_{H}\to Z^0H^0)$,
where $f=u,d,c,s,b,t,$ $e,\mu,\tau,\nu_e,\nu_{\mu},\nu_{\tau}$.

\vskip 5mm
\par
\section{Analytical calculations }
\subsection{ LO calculations }
\par
\begin{figure*}
\begin{center}
\includegraphics*[scale=1]{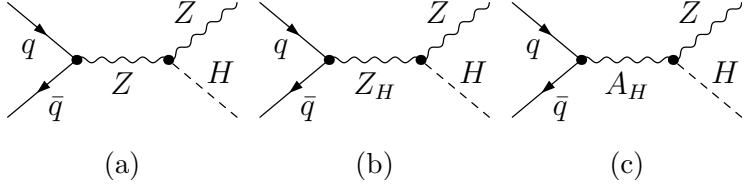}
\caption{\label{fig1} The LO Feynman diagrams for the
\qqzh($q=u,d,s,c,b$) partonic process. }
\end{center}
\end{figure*}

We generate the Feynman diagrams and their corresponding amplitudes
by using FeynArts3.5 package \cite{fey}, and apply FormCalc5.4
package \cite{formcalc} to implemented the amplitude simplification.
The LO contribution to the \ppzh process comes from $q$-$\bar{q}$
annihilation $(q=u,d,c,s,b)$. We denote the partonic process as
\begin{equation}
q(p_{1}) + \bar q (p_{2})\to Z^0(p_{3}) + H^0(p_{4}),\hspace{0.5cm}
(q=u,d,c,s,b),
\end{equation}
where $p_{1}$, $p_{2}$, $p_{3}$ and $p_{4}$ represent the
four-momenta of incoming partons, the outgoing $Z^0$- and
$H^0$-boson, respectively. We use the 't Hooft-Feynman gauge
throughout our calculations. Comparing with the partonic colliding
energy at the LHC, the quark masses, $m_{q}~(q=u,d,c,s,b)$, are
relatively small. We neglect their masses in our further
calculations. The Feynman diagram for the \qqzh subprocess in the SM is
shown in Fig.\ref{fig1}(a). The amplitudes corresponding to
Figs.\ref{fig1}(a), (b) and (c) without introducing the decay widths
in propagators are denoted as ${\cal M}_{LO}^{Z}(\Gamma_Z=0)$,
${\cal M}_{LO}^{Z_H}(\Gamma_{Z_H}=0)$ and ${\cal
M}_{LO}^{A_H}(\Gamma_{A_H}=0)$, respectively.

\par
As shown in Eq.(\ref{Z-coupling-1})-Eq.(\ref{Z-coupling-4}), the
coupling between $Z^0$ and quarks in the LHM can be obtained from
the SM one with a correction of ${\cal O}(v^2/f^2)$. The \qqzh
subprocess in the LHM obtains additional contributions coming from
the diagrams with the exchange of heavy gauge bosons $Z_H$ and $A_H$
shown in Fig.\ref{fig1}(b)-(c). These two heavy neutral gauge
bosons, $Z_H$ and $A_H$, are potentially resonant. For disposal of
the singularities due to $V_H$ $(V_H=Z_H,A_H)$ resonances in the
calculations, we have to introduce the decay widths of $Z_H$ and
$A_H$ by doing the following replacements in the resonance
propagators of the amplitudes ${\cal M}^{Z_H}_{LO}(\Gamma_{Z_H}=0)$
for Fig.\ref{fig1}(b) and ${\cal M}^{A_H}_{LO}(\Gamma_{A_H}=0)$ for
Fig.\ref{fig1}(c),
\begin{eqnarray}
\frac{1}{\hat{s}_{12}-M_{V_H}^2} \to \frac{1}{\hat{s}_{12}-M_{V_H}^2+i
M_{V_H}\Gamma_{V_H}}, \label{Replace}
\end{eqnarray}
where $\Gamma_{V_H}$ $(V_H=Z_H,A_H)$ represents the decay width of
$V_H$. Then we get the LO amplitudes for Fig.\ref{fig1}(b) and (c)
at the tree-level respectively expressed as
\begin{eqnarray}
{\cal M}_{LO}^{Z_H}&=&\frac{\hat{s}_{12}-M_{Z_H}^2}{\hat{s}_{12}-M_{Z_H}^2+i
{\cal M}_{Z_H}\Gamma_{Z_H}}{\cal M}_{LO}^{Z_H}(\Gamma_{Z_H}=0), \nb \\
{\cal M}_{LO}^{A_H}&=&\frac{\hat{s}_{12}-M_{A_H}^2}{\hat{s}_{12}-M_{A_H}^2+i
{\cal M}_{A_H}\Gamma_{A_H}}{\cal M}_{LO}^{A_H}(\Gamma_{A_H}=0).
\end{eqnarray}
The modified amplitudes ${\cal M}_{LO}^{Z_H}$ and ${\cal
M}_{LO}^{A_H}$ are safe amplitudes being free of the $Z_H$ and $A_H$
resonance singularities. Since the ${\cal O}(\alpha_s)$ corrections
do not contribute to the LO $Z_H$ and $A_H$ decay widths, these
replacements cannot induce the double-counting problem in our NLO
calculations.

\par
The LO cross section for the subprocess \qqzh is expressed as
\begin{equation}
\label{LO}\hat{\sigma}_{LO}^{q\bar
q}=\frac{1}{4}\frac{1}{9}\frac{(2\pi)^{4}}{2  {\hat s}^{2}} \int
\sum_{spin}^{color}|{\cal M}_{LO}|^{2}d\Omega_{2},~~(q=u,d,c,s,b)
\end{equation}
where the factors $\frac{1}{4}$ and $\frac{1}{9}$ come from the
averaging over the spins and colors of the initial partons
respectively, $\hat s$ is the partonic center-of-mass energy
squared, and ${\cal M}_{LO}$ is the amplitude of all the LO diagrams
shown in Fig.\ref{fig1}. The summation is taken over the spins and
colors of all the relevant particles in the \qqzh subprocess. The
integration is performed over the two-body phase space of the final
particles $Z^0$ and $H^0$. $d\Omega_{2}$ is the two-body phase space
element expressed as
\begin{eqnarray}\label{PhaseSpace}
{d\Omega_{2}}=\delta^{(4)} \left( p_1+p_2-\sum_{i=3}^4 p_i \right)
\prod_{j=3}^4 \frac{d^3 \textbf{\textsl{p}}_j}{(2 \pi)^3 2 E_j}.
\end{eqnarray}

\par
Within the framework of the QCD factorization, the LO cross section
for the process \ppzh at the LHC can be obtained by performing the
following integration of the cross section for the subprocess \qqzh
over the partonic luminosities (see Eq.(\ref{integration})).
\begin{equation}
\label{integration} \sigma_{LO}= \sum_{ij=u\bar u,d\bar d,}^{s\bar
s,c\bar c,b\bar b,} \int_{0}^{1}dx_1 \int_{0}^{1} dx_2 \left[
G_{i/P_1}(x_1,\mu_f) G_{j/P_2}(x_2,\mu_f)+(x_1 \leftrightarrow x_2,
P_1 \leftrightarrow P_2)\right] \hat{\sigma}_{LO}^{i j}(\hat{s}=x_1
x_2 s),
\end{equation}
where $G_{i/A}(x,\mu_f)$ ($i=u,d,s,c,b$) is parton distribution
function (PDF) of proton $A~(=P_1,P_2)$ which describes the
probability to find a parton $i$ with momentum $xp_A$ in proton $A$,
$s$ is defined as the total colliding energy squared in
proton-proton collision, $\hat{s}=x_1x_2 s$, and $\mu_f$ is the
factorization scale. In our LO calculations, we adopt the CTEQ6L1
\cite{pdfs} PDFs.

\vskip 5mm
\par
\subsection{ Virtual and real emission corrections }
\par
\begin{figure*}
\begin{center}
\includegraphics*[scale=1]{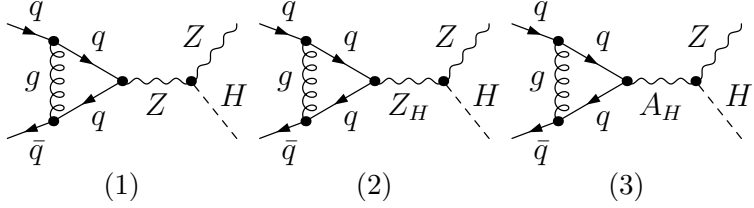}
\caption{\label{fig2} The QCD one-loop vertex correction diagrams
for the subprocess \qqzh ($q\bar q= u\bar u, d\bar d, c\bar c, s\bar
s, b\bar b$). }
\end{center}
\end{figure*}

The QCD one-loop vertex correction diagrams for the partonic process
\qqzh with nonzero contribution are presented in Fig.\ref{fig2}.
There exist both ultraviolate (UV) and soft/collinear infrared (IR)
singularities in the one-loop diagrams. We
regularize all the singularities by using the dimensional
regularization method in $D=4-2\epsilon$ dimensions, and apply the
modified minimal subtraction ($\overline{\rm MS}$) scheme to
renormalize the relevant fields. The UV divergence of the virtual
corrections are removed by renormalized wave functions of the
relevant quarks. We define the renormalization constants of the
relevant quark fields as
\begin{eqnarray}
\label{defination} \psi_{q}^{0,L,R} &=& \left( 1 + \frac{1}{2}
\delta Z_{q}^{L,R} \right) \psi_{q}^{L,R},
\end{eqnarray}
where $\psi^{L,R}_{q}$ denotes the field of the SM quark. Its
renormalization constant are expressed as
\begin{eqnarray} \label{CT-q}
\delta Z_{q} & \equiv & \delta Z_{q}^{L}=\delta Z_{q}^{R}=
-\frac{\alpha_s(\mu_r)}{3 \pi} \Big[ \Delta_{UV}-\Delta_{IR} \Big].
\end{eqnarray}
The notations used in above equation are defined as
$\Delta_{UV}=1/\epsilon_{UV} -\gamma_E +\ln(4\pi)$ and
$\Delta_{IR}=1/\epsilon_{IR} -\gamma_E +\ln(4\pi)$.

\par
\begin{figure*}
\begin{center}
\includegraphics*[scale=1]{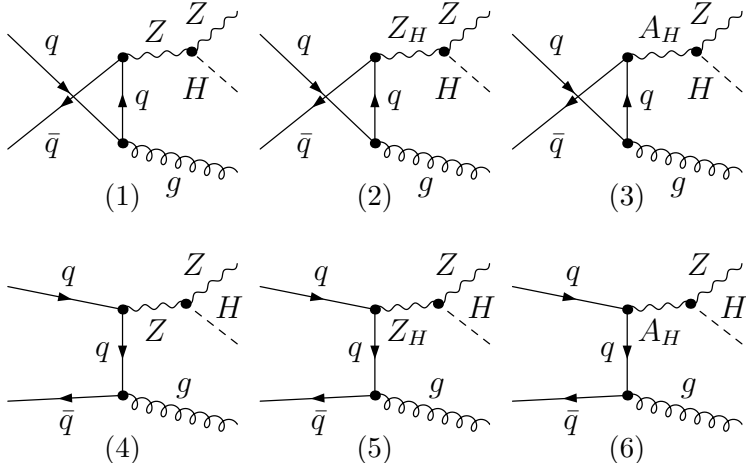}
\caption{\label{fig3} The Feynman diagrams for the real
gluon emission subprocess \qqzhg$ (q=u,d,c,s,b)$. }
\end{center}
\end{figure*}
\begin{figure*}
\begin{center}
\includegraphics*[scale=1]{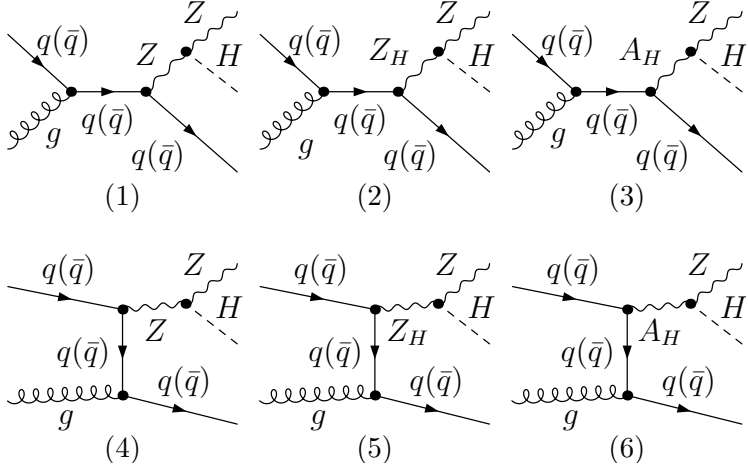}
\caption{\label{fig4} The Feynman diagrams for the real
light-quark emission subprocess \qgzhq$(q=u,d,c,s,b)$.}
\end{center}
\end{figure*}

Although the total NLO QCD amplitude of subprocess \qqzh is UV
finite after performing renormalization procedure, it still contains
soft/collinear IR singularities. The soft IR singularity can be
completely canceled by the contribution of real gluon emission
subprocess \qqzhg, while the collinear singularity is eliminated
partially by the light-quark emission subprocesses \qgzhq. The
remaining collinear IR divergence can be absorbed by the
counterterms of PDFs. We adopt the analytical expressions for
IR-singular parts of loop integrals from Ref.\cite{Stefan}, and use
the expressions in Refs.\cite{OneTwoThree,Four,Five} to implement
the numerical evaluations of IR-safe $N$-point($N\leq4$) integrals.
The Feynman diagrams for real gluon/light-quark eimission are
depicted in Fig.\ref{fig3} and Fig.\ref{fig4}, respectively.

\par
We apply the two cutoff phase space slicing (TCPSS) method
\cite{TCPSS} to isolate the soft and collinear IR singularities of
the real emission correction from the IR-safe region. In performing
the calculations with the TCPSS method, we should introduce
arbitrary small soft cutoff $\delta _{s}$ and collinear cutoff
$\delta _{c}$. The phase space of the $q(p_1) \bar q(p_2) \to
Z^0(p_3)H^0(p_4)g(p_5)$ partonic process can be split into two
regions, $E_5 \leq \delta_s\sqrt{\hat{s}}/2$ (soft gluon region) and
$E_5 > \delta_s\sqrt{\hat{s}}/2$ (hard gluon region) by soft cutoff
$\delta_s$. The hard gluon region is separated as hard collinear
($\rm HC$) and hard non-collinear ($\overline {\rm HC}$) regions by
cutoff $\delta_c$. The ${\rm HC}$ region is the phase space where
$-\hat{t}_{15}$(or $-\hat{t}_{25}$)$<\delta_c \hat{s}$
$(\hat{t}_{15}\equiv(p_1-p_5)^2$ and
$\hat{t}_{25}\equiv(p_2-p_5)^2)$. The phase space of light-quark
emission $q(p_1)[(\bar q(p_1)]g(p_2) \to Z^0(p_3)H^0(p_4)q(p_5)[\bar
q(p_5)]$ is split into hard collinear ($\rm HC$) region and hard
non-collinear ($\overline {\rm HC}$) region by introducing a cutoff
$\delta_c$. The real gluon emission corrections over the $\overline
{\rm HC}$ region are finite and can be calculated numerically with
general Monte Carlo method \cite{Lepage}. Finally, the cross section
for the real emission partonic process can be written as
\begin{equation}
\label{sigmaR}
\hat{\sigma}_{R}=\hat{\sigma}_{S}+\hat{\sigma}_{H}
=\hat{\sigma}_{S}+\hat{\sigma}_{HC}+\hat{\sigma}_{\overline{HC}}.
\end{equation}

\par
With the NLO correction components given above, the full QCD NLO
corrected cross section for the $Z^0H^0$ production at the
LHC can be formally obtained by the QCD factorization formula as
\begin{eqnarray}
&& \sigma_{NLO}(pp \to Z^0H^0 +X) = \nonumber \\
&& \int dx_{P_1} dx_{P_2} \left\{ \sum_{ij}\left[
G_{i/P_1}(x_{P_1},\mu_f) G_{j/P_{2}}(x_{P_2},\mu_f)
\hat{\sigma}^{ij}_{NLO}(x_{P_1}x_{P_2}s,\mu_r)\right]
+(P_1\leftrightarrow P_2)\right\}, \nb \\
\end{eqnarray}
where $i$ and $j$ run over all possible initial partons contributing
to the \ppzh process up to the QCD NLO, and the notations of
$\mu_f$, $x_{P_1}$, $x_{P_2}$ are the same as in
Eq.(\ref{integration}). We adopt the CTEQ6m PDFs for
$G_{i/P_{1}}(x_{P_1},\mu_f)$ and $ G_{j/P_{2}}(x_{P_2},\mu_f)$ in
the NLO calculations \cite{pdfs}. The total QCD NLO corrected cross
section for partonic process \qqzh can be expressed as
\begin{eqnarray}
\hat\sigma^{ij}_{NLO}=\hat\sigma^{ij}_{LO}
+\Delta\hat\sigma^{ij}_{NLO}=\hat\sigma^{ij}_{LO}+
\hat{\sigma}^{ij}_{R}+\hat{\sigma}^{ij}_{V},
\end{eqnarray}
where $\hat {\sigma}^{ij}_{LO}$,$\hat{\sigma}_{R}^{ij}$ and
$\hat{\sigma}^{ij}_{V}$ denote the cross sections for tree level,
real emission and virtual corrections for parton level process,
respectively.

\vskip 5mm
\par
\section{Numerical results and discussions}
\par
In this section we provide and discuss the numerical results for the
\ppzh process in the LHM up to the QCD NLO. In order to make a cross
check with previous work on the $Z^0H^0$ associated production in
the SM at the LHC, we take $\mu=\mu_f=\mu_r=\sqrt{s_{ZH}}$, and the
input parameters and PDFs being the same as used in
Ref.\cite{smwork}, and calculate the LO and NLO QCD corrected total
cross sections for \ppzh at the $\sqrt{s}= 14~TeV$ LHC in the SM. We
get the total cross sections for $M_H=140~GeV$ as
$\sigma_{LO}=0.46827(3)~pb$ and $\sigma_{NLO}=0.5770(4)~pb$,
separately. The corresponding results can be read out from Table 8
of Ref.\cite{smwork}: $\sigma_{LO}=0.4684(2)~pb$ and
$\sigma_{NLO}=0.5768(2)~pb$, which are coincident with ours within
the calculation errors.

\par
In our following numerical calculations we take the colliding energy
in proton-proton center-of-mass system as $\sqrt{s}=8~TeV$ for the
early LHC and $\sqrt{s}=14~TeV$ for the future LHC. We use one- and
two-loop running $\alpha_{s}(\mu)$ by taking $\Lambda_5^{LO} =
165~MeV$ and $\Lambda_5^{\overline{MS}} = 226~MeV$ for the LO and
NLO calculations, respectively \cite{hdata}. The factorization and
the renormalization scales are set to be equal for simplicity
($\mu\equiv\mu_f = \mu_r$). We take $\mu=\mu_0= (M_H+M_Z)/2$ in
default unless otherwise stated. We neglect the masses of $u$-,
$d$-, $c$-, $s$-, and $b$-quark, and take
\begin{eqnarray}
&& G_{\mu}= 1.16637\times 10^{-5}~GeV^{-2},~~~~
M_W=80.399~GeV, \nb  \\
&& M_Z=91.1876~GeV,~~~~m_t=171.2~GeV,~~~~ M_H=125~GeV.
\end{eqnarray}
The $G_{\mu}$ scheme is adopted, i.e., the electromagnetic coupling
$\alpha$ is derived from $\alpha_{G_{\mu}}= \sqrt{2}G_{\mu}
M_W^2\left(1-M_W^2/M_Z^2\right)/\pi$. Considering the constraints of
the electroweak precision data on LHM parameters \cite{range-1}, we
assumed that $0.1 < c <0.5$, $0.1 < c^{\prime} <0.9$ and $3~TeV < f
< 7~TeV$, and take the representative input LHM parameter set as
$c=0.5$, $c^{\prime}=0.22$ and $f=4~TeV$ in our numerical
calculations if there is no other statement. From
Eqs.(\ref{WH-mass}), (\ref{AH-mass}) and (\ref{ZH-mass}) with this
input parameter set the masses of the heavy gauge bosons $M_{A_H}$,
$M_{W^{\pm}_H}$ and $M_{Z_H}$ are obtained as $1.461~TeV$,
$3.025~TeV$ and $3.025~TeV$ respectively, where the mass values of
heavy vector gauge bosons $Z^0_H$ and $W^\pm_H$ are beyond the
corresponding experimental lower mass limits \cite{CDF-ZH}.

\par
\begin{figure}[htbp]
\includegraphics[scale=0.5]{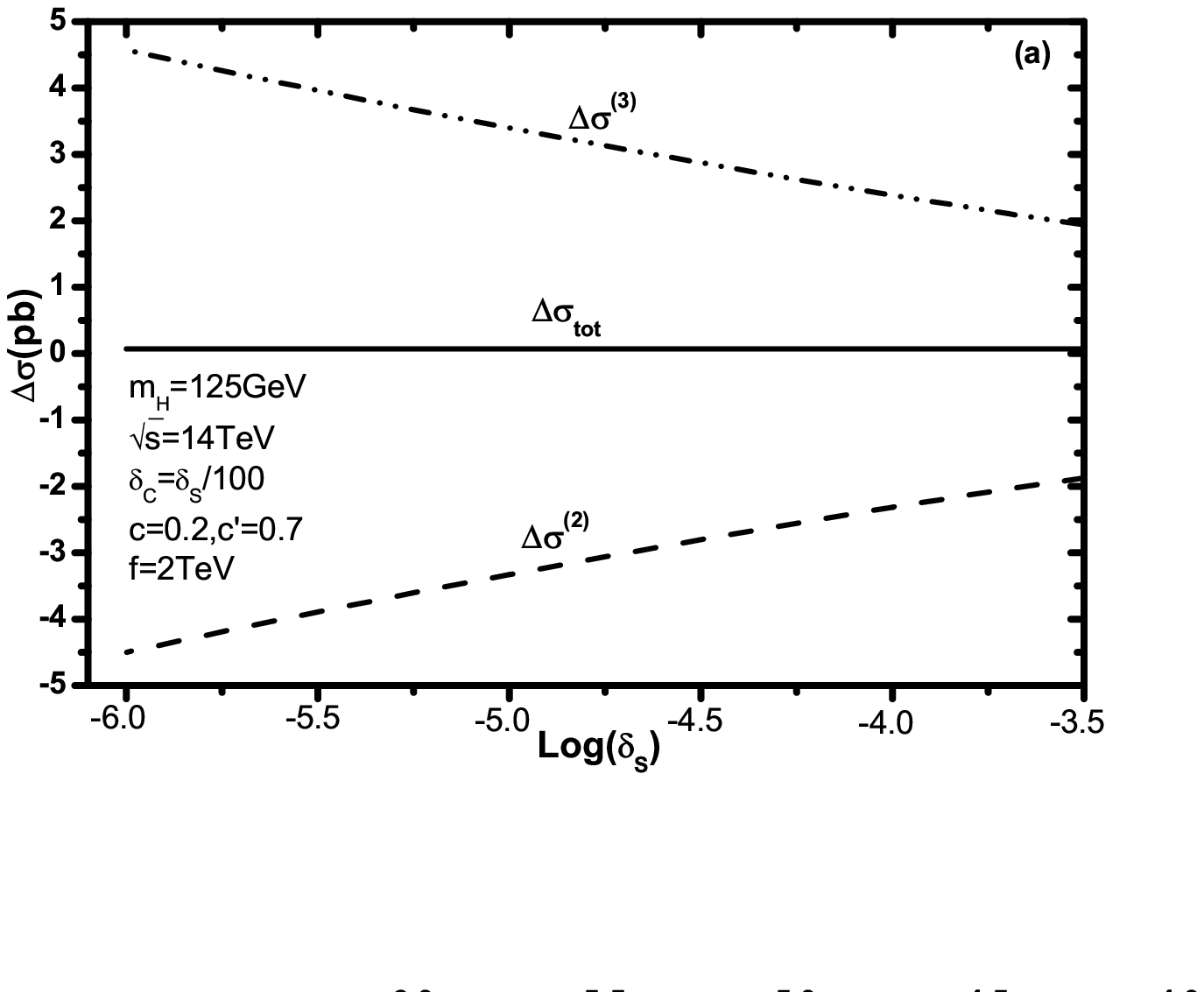}
\includegraphics[scale=0.5]{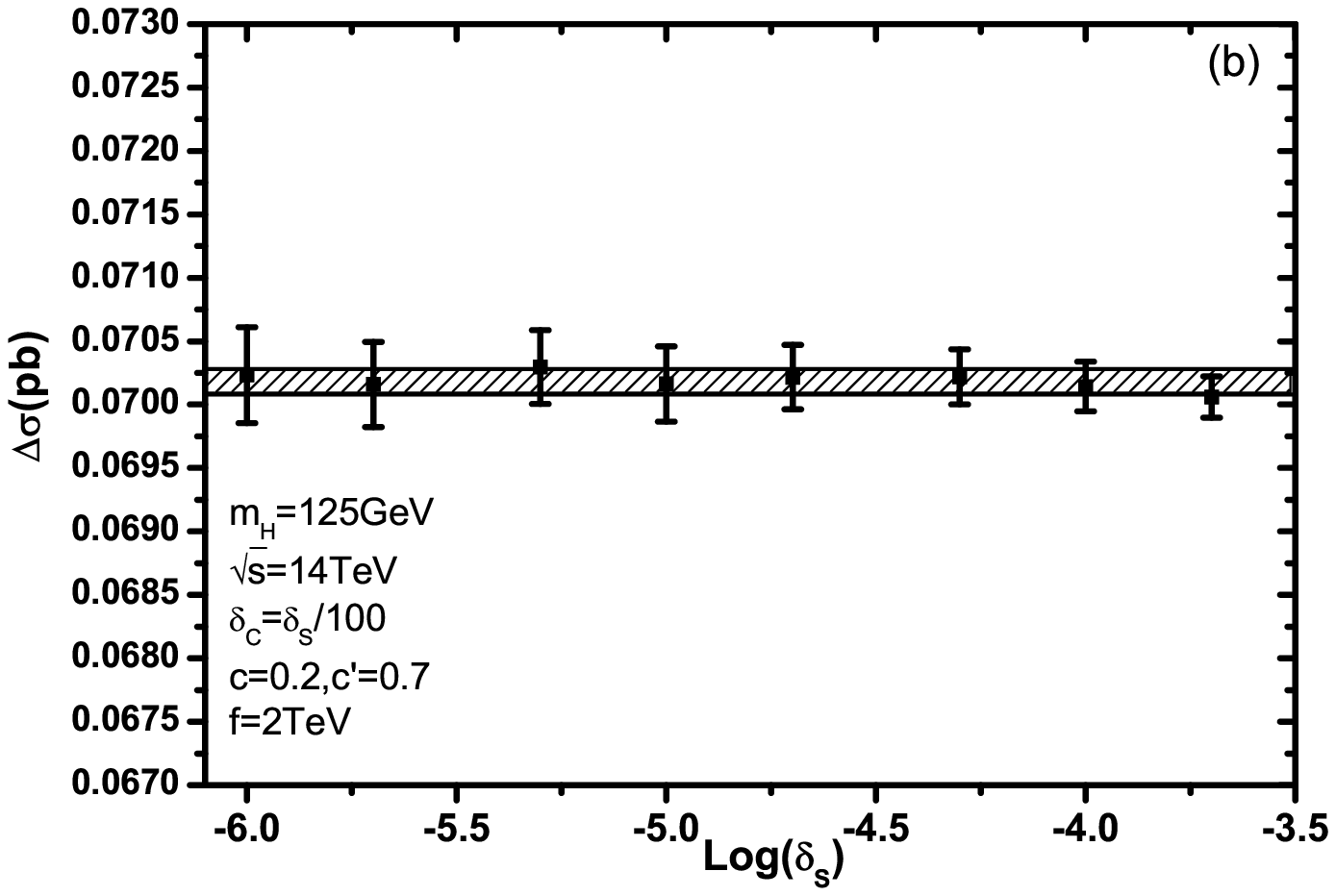}
\hspace{0in}%
\caption{\label{fig5} (a) The NLO QCD corrections to the \ppuuzh
process in the LHM at the $\sqrt{s}=14~TeV$ LHC as the functions of
the soft cutoff $\delta_s$, where we take $c=0.2$, $c^{\prime}=0.7$
and $f=2~TeV$, $\delta_c=\delta_s/100$ and $\mu=\mu_0$. (b) The
amplified curve for the NLO QCD correction to the cross section
$\Delta \sigma_{NLO}$.}
\end{figure}

In order to verify the independence of the total NLO QCD corrections
on the introduced arbitrary cutoff values of $\delta_s$ ($\delta
_c$), we depict the $\Delta\sigma_{NLO}$ for the \ppuuzh process in
the LHM at the $\sqrt{s}=14~TeV$ LHC as the functions of $\delta_s$
in Figs.\ref{fig5}(a,b), where we take $c=0.2$, $c^\prime=0.7$,
$f=2~TeV$, $\delta_c=\delta_s/100$ and $\mu=\mu_0$. The amplified
curve for the total NLO QCD correction ($\Delta \sigma_{NLO}$) for
the process \ppuuzh is shown in Fig.\ref{fig5}(b). We can see in
Figs.\ref{fig5}(a,b) that the total QCD correction to the \ppzh
process does not depend on the arbitrarily chosen value of the
cutoffs $\delta_s$ and $\delta_c$. The two-body correction ($\Delta
\sigma ^{(2)}$) and three-body correction ($\Delta \sigma ^{(3)}$)
and the total QCD correction ($\Delta \sigma_{NLO}=\Delta \sigma
^{(2)}+\Delta \sigma ^{(3)}$) for the \ppuuzh process at the LHC are
depicted as the functions of the soft cutoff $\delta_s$ in
Figs.\ref{fig5}(a). The curve for $\Delta \sigma_{NLO}$ is presented
in Fig.\ref{fig5}(b) together with calculation errors. We adopt also
the dipole subtraction (DPS) method to deal with the IR
singularities for further verification. The $\Delta \sigma_{NLO}$
results from the DPS method including $\pm 1\sigma$ statistic errors
are plotted as the shadowing region in Fig.\ref{fig5}(b). It shows
that the results by using both the TCPSS method and the DPS method
are in good agreement. In further numerical calculations we adopt
the TCPSS method and fix $\delta_s=1\times10^{-5}$ and
$\delta_c=1\times10^{-7}$.

\par
\begin{figure}[htbp]
\includegraphics[scale=0.45]{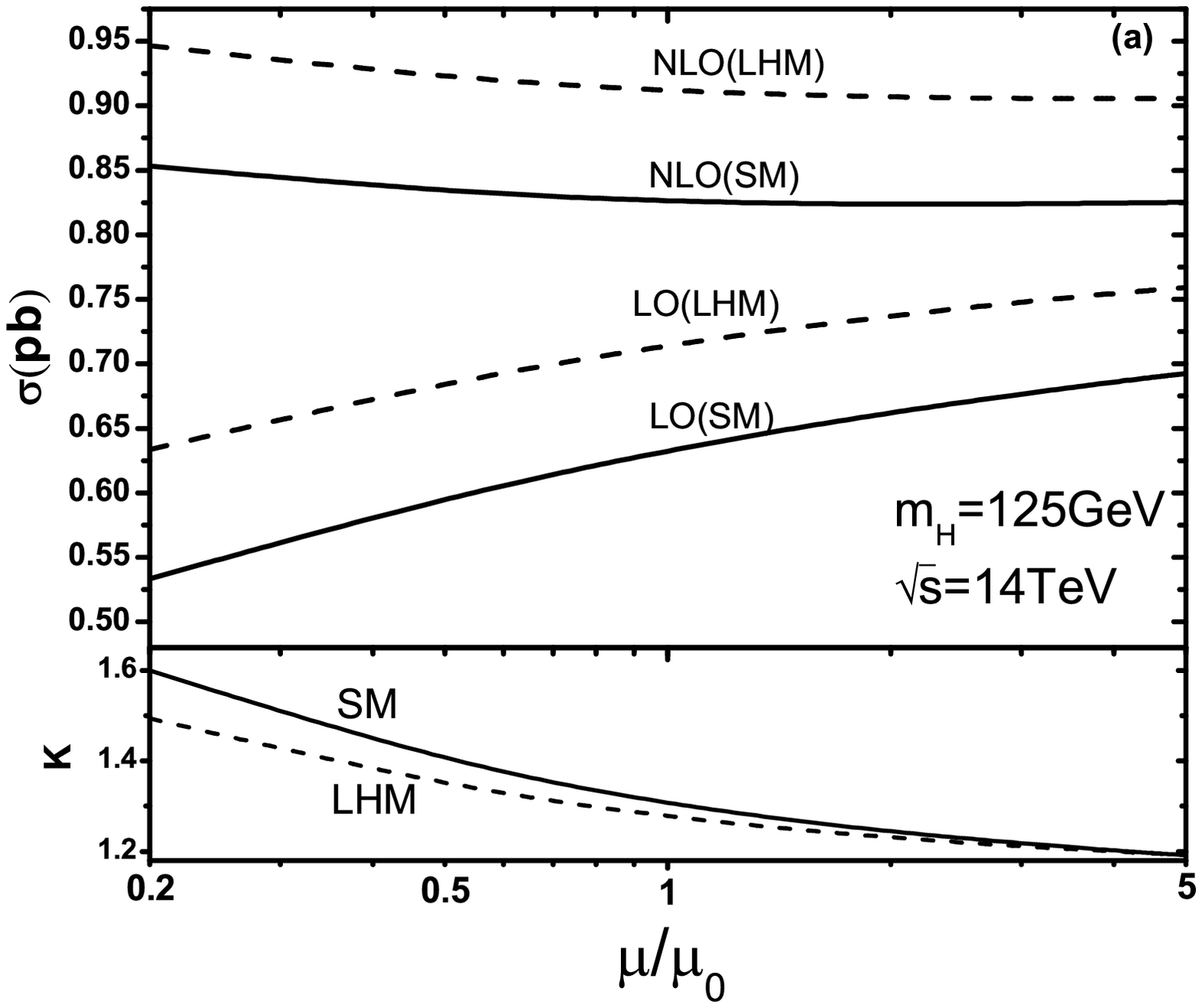}
\includegraphics[scale=0.45]{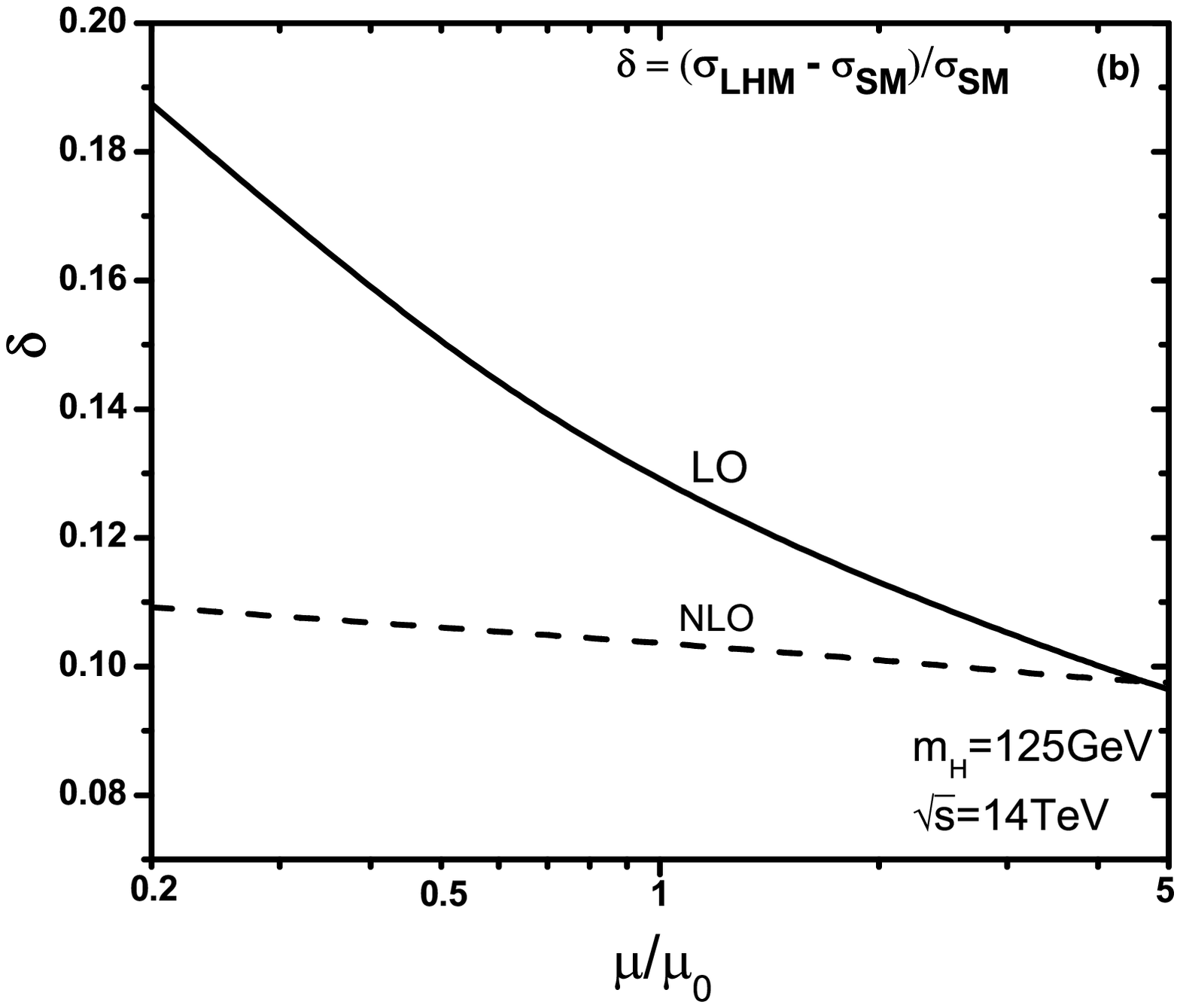}
\includegraphics[scale=0.45]{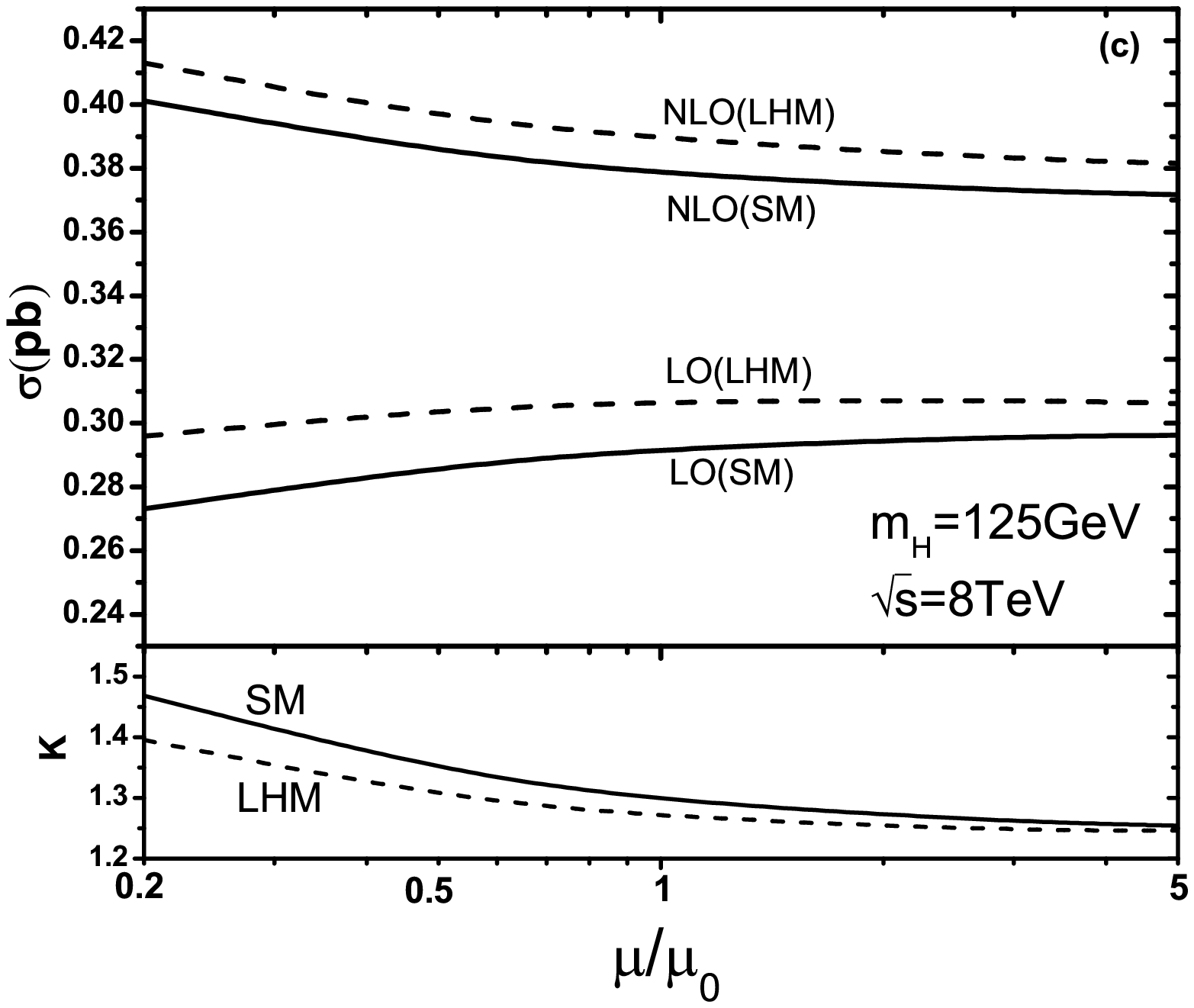}
\includegraphics[scale=0.45]{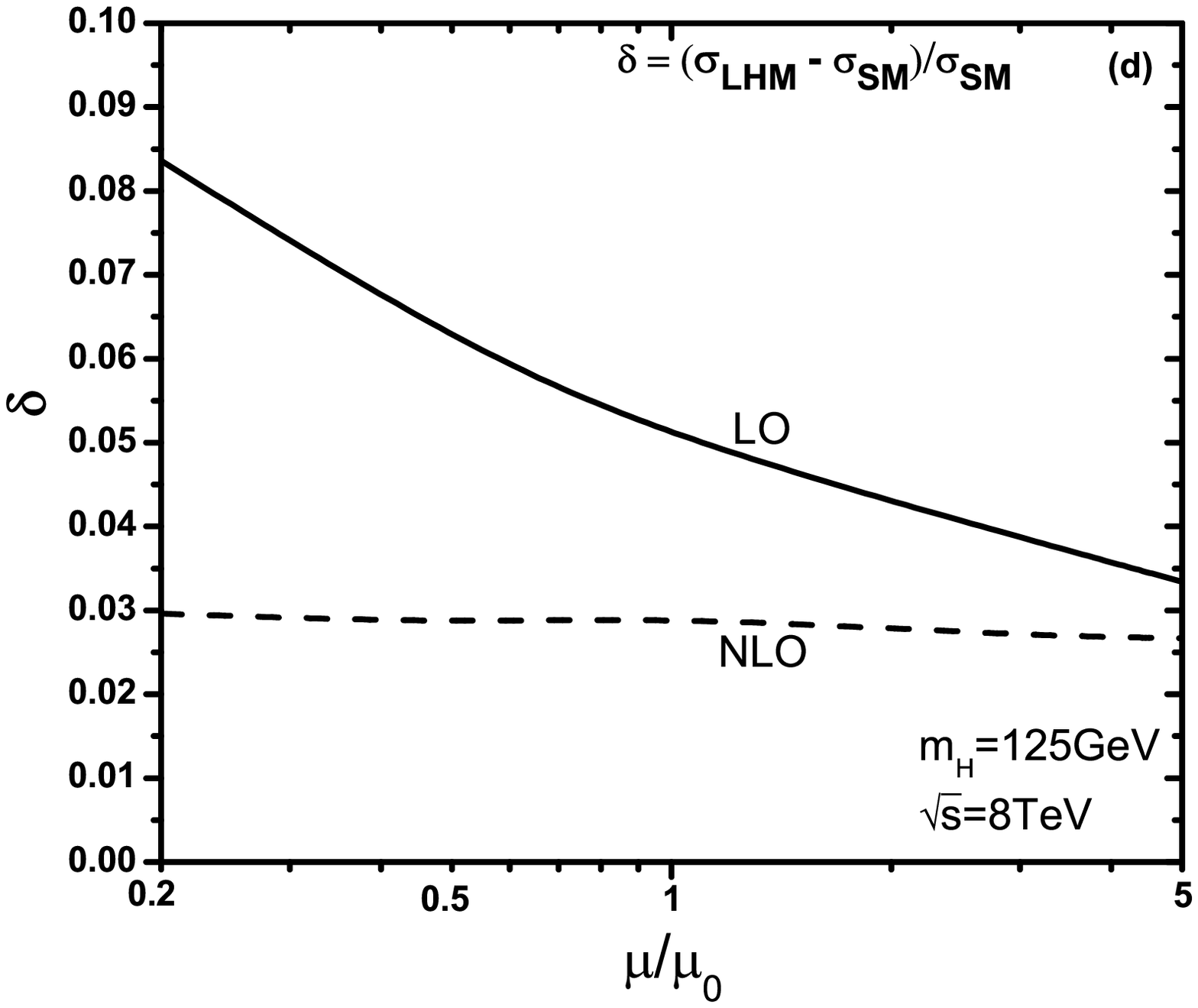}
\hspace{0in}%
\caption{\label{fig6} In these four plots we take $c=0.5$,
$c^{\prime}=0.22$ and $f=4~TeV$. (a) The dependence of the LO and
the QCD corrected cross sections and the corresponding K-factor for
the process \ppzh on the factorization/renormalization scale
($\mu/\mu_0$) at the $\sqrt{s}=14~TeV$ LHC. (b) The corresponding
relative deviation of the integrated cross sections in the LHM from
those in the SM, as the functions of $\mu/\mu_0$ at the
$\sqrt{s}=14~TeV$ LHC. (c) The LO and the QCD corrected cross
sections and the corresponding K-factor for the process \ppzh versus
the scale $\mu/\mu_0$ at the $\sqrt{s}=8~TeV$ LHC. (d) The
corresponding relative deviation of the integrated cross sections in
the LHM from those in the SM, as the functions of $\mu/\mu_0$ at the
$\sqrt{s}=8~TeV$ LHC. }
\end{figure}

We show the integrated LO, NLO QCD corrected cross sections and the
corresponding K-factor ($K(\mu)\equiv
\sigma_{NLO}(\mu)/\sigma_{LO}(\mu)$) at the $\sqrt{s}=14~TeV$
($\sqrt{s}=8~TeV$) LHC for the process \ppzh as the functions of the
factorization/renormalization scale ($\mu/\mu_0$) in
Figs.\ref{fig6}(a) (Figs.\ref{fig6}(c)), where we set $\mu \equiv
\mu_r=\mu_f$, $\mu_0\equiv (M_H+M_Z)/2$, $c=0.5$, $c^{\prime}=0.22$
and $f=4~TeV$. If we define the scale uncertainty for the \ppzh
process as $\eta = \frac{|\sigma
(\mu=5\mu_0)-\sigma(\mu=0.2\mu_0)|}{\sigma(\mu=\mu_0)}$, from the
curves in Figs.\ref{fig6}(a,c) we can figure out the corresponding
uncertainties at the $\sqrt{s}=14~TeV$ LHC being
$\eta^{SM}_{LO}=0.251$, $\eta^{SM}_{NLO}=0.034$,
$\eta^{LHM}_{LO}=0.176$ and $\eta^{LHM}_{NLO}=0.045$, and at the
$\sqrt{s}=8~TeV$ LHC $\eta^{SM}_{LO}=0.080$,
$\eta^{SM}_{NLO}=0.081$, $\eta^{LHM}_{LO}=0.033$ and
$\eta^{LHM}_{NLO}=0.081$, respectively. We can see that at the
$\sqrt{s}=14~TeV$ LHC the LO cross sections are strongly related to
the scale in the plotted $\mu$ range, and the NLO QCD corrections
significantly reduce the scale uncertainties. But at the
$\sqrt{s}=8~TeV$ LHC there is no distinct improvement for the scale
dependence when the QCD NLO corrections are involved.
Fig.\ref{fig6}(b) and Fig.\ref{fig6}(d) present the relative
deviations defined as $\delta(\mu)\equiv
\frac{\left[\sigma^{LHM}(\mu)-\sigma^{SM}(\mu)\right]}
{\sigma^{SM}(\mu)}$, as the functions of $\mu/\mu_0$, which
correspond to Fig.\ref{fig6}(a) and Fig.\ref{fig6}(c), respectively.
The two figures demonstrate that the NLO QCD corrections obviously
reduce the relative deviation $\delta$ in our plotted $\mu/\mu_0$
range. The theoretical relative deviations including the NLO
corrections are above $9.75\%$ and $2.67\%$ at the $\sqrt{s}=14~TeV$
and $\sqrt{s}=8~TeV$ LHC, respectively. We can read out from the
figures that the effects ($\delta$) from the heavy neutral gauge
boson interactions at the $\sqrt{s}=14~TeV$ LHC in the vicinity of
$\mu=\mu_0$ can be about $12.83\%$ for $\sigma_{LO}$ and $10.37\%$
for $\sigma_{NLO}$.

\par
\begin{figure}[htbp]
\includegraphics[scale=0.45]{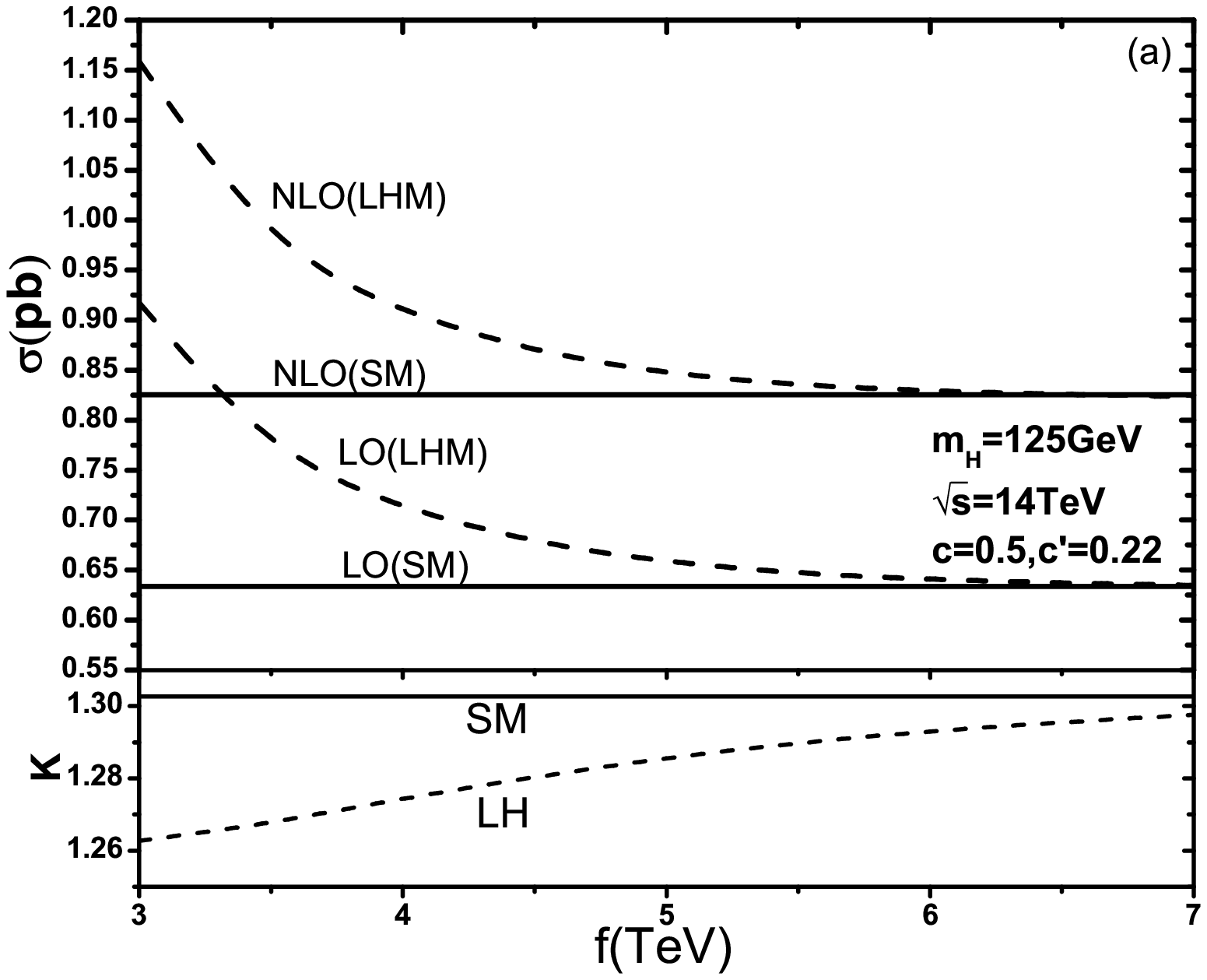}
\includegraphics[scale=0.45]{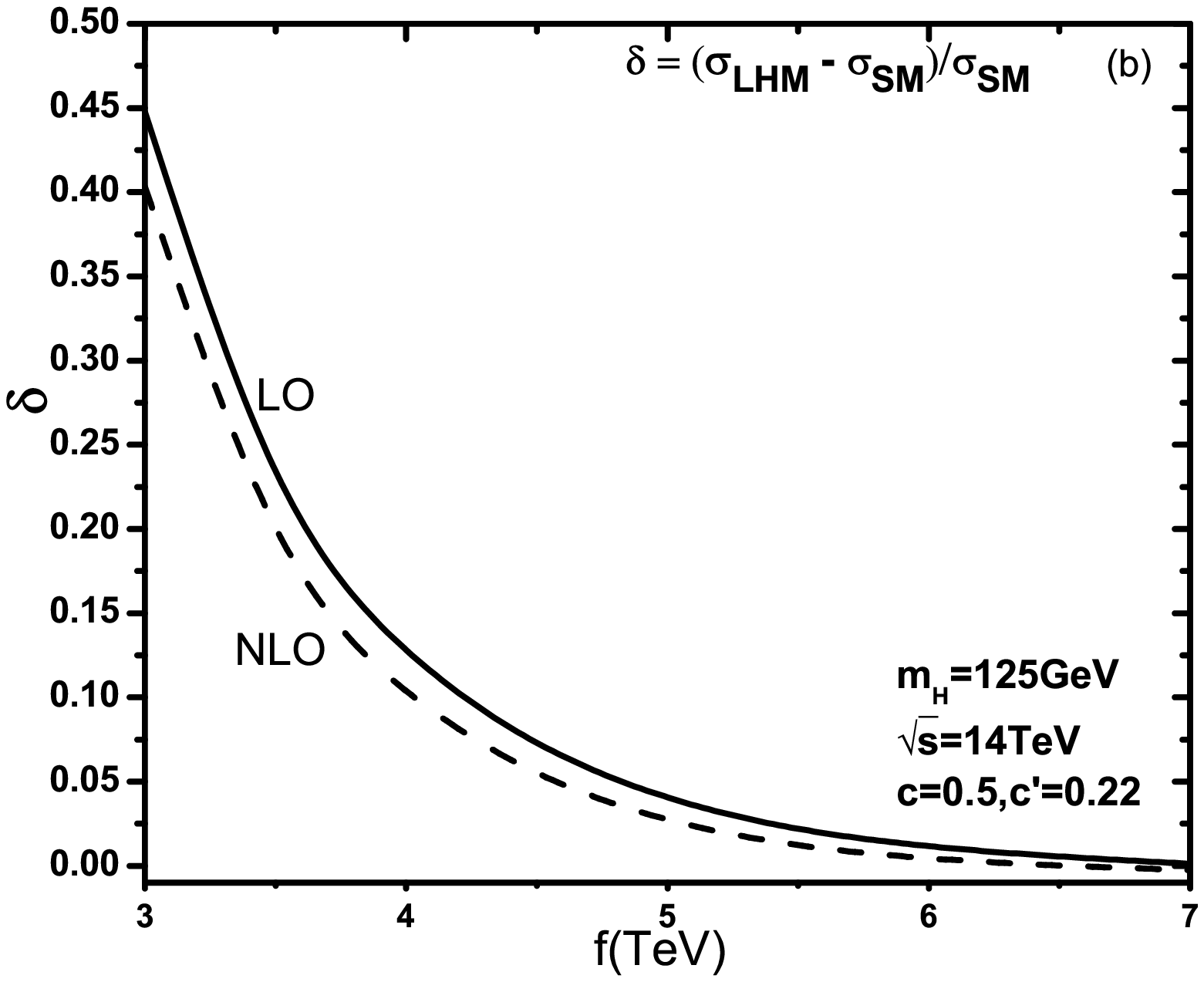}
\includegraphics[scale=0.45]{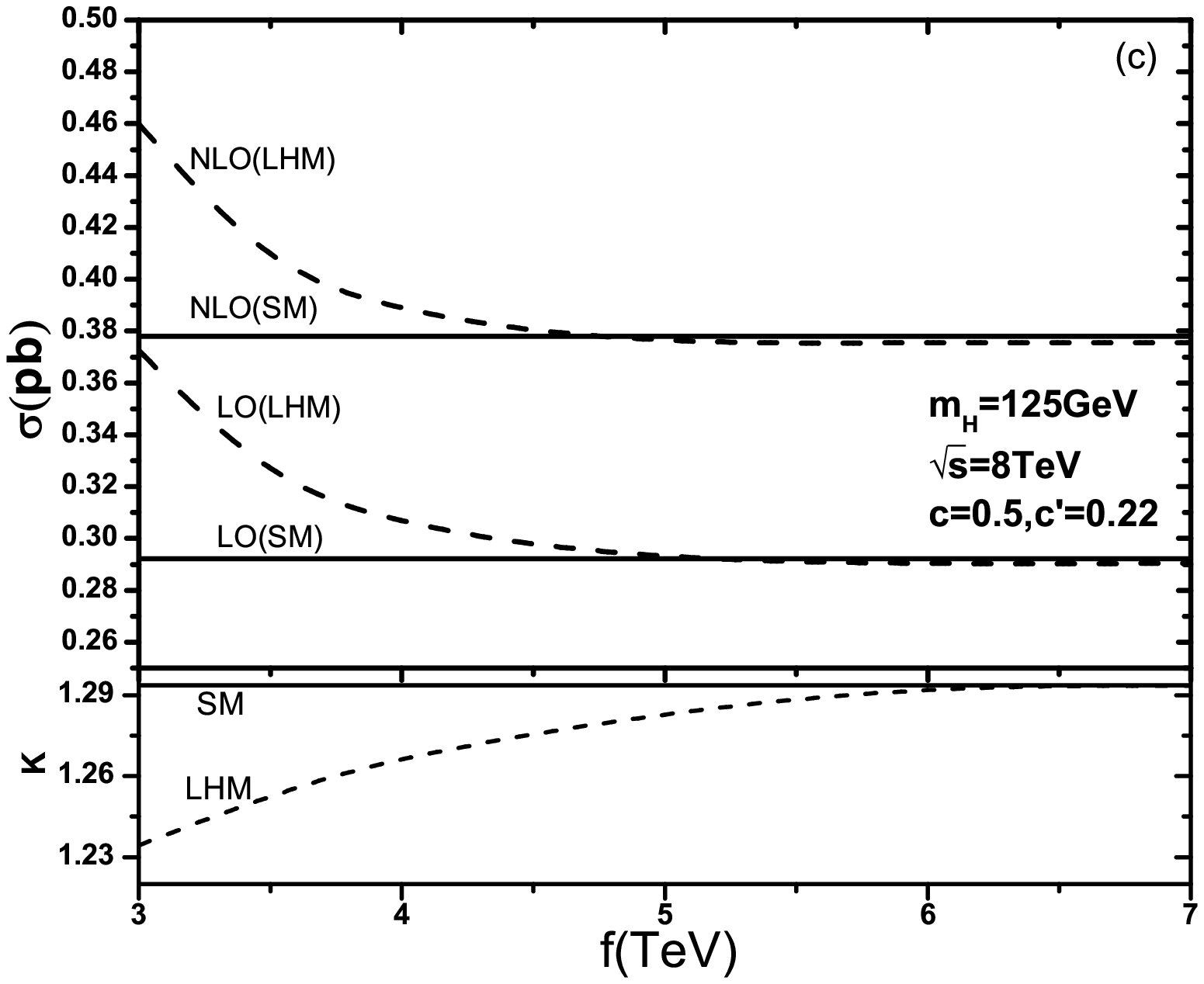}
\includegraphics[scale=0.45]{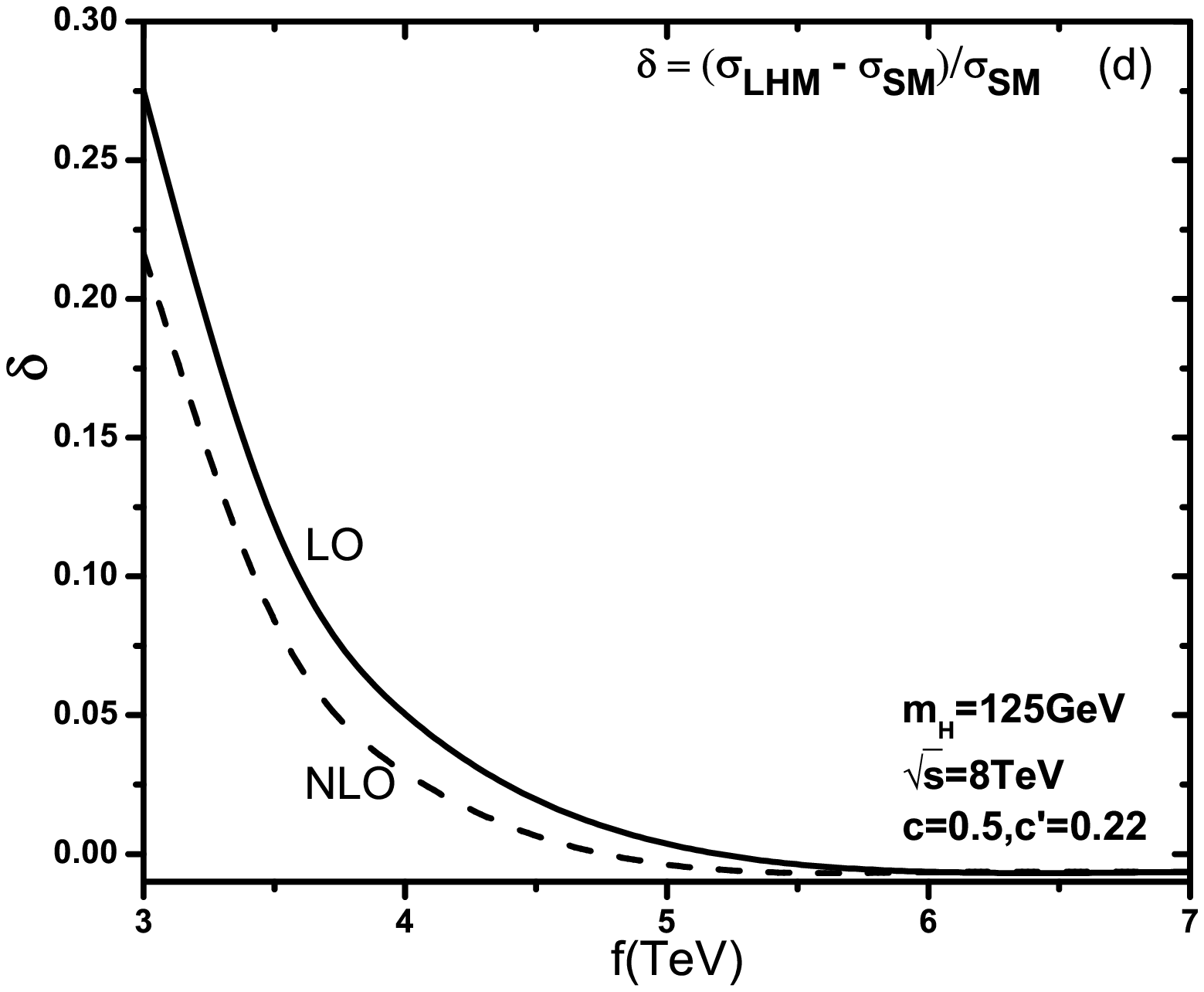}
\hspace{0in}%
\caption{\label{fig7} We take $c=0.5$, $c^{\prime}=0.22$ and
$\mu=\mu_0$. (a) The LO and NLO QCD corrected cross sections and the
corresponding K-factors for the $pp \to Z^0H^0+X$ process at the
$\sqrt{s}=14~TeV$ LHC in both the SM and LHM as the functions of
scale $f$. (b) The relative deviations of the cross sections in the
LHM from those in the SM corresponding to Fig.\ref{fig7}(a) as the
functions of scale $f$. (c) The LO and NLO QCD corrected cross
sections at the $\sqrt{s}=8~TeV$ LHC in both the SM and the LHM as
the functions of scale $f$. (d) The relative deviations
corresponding to Fig.\ref{fig7}(c) as the functions of scale $f$. }
\end{figure}

In following analysis we show the influence of the LHM parameters
$c$, $c^\prime$, and the global symmetry breaking scale $f$. In
Figs.\ref{fig7}(a,b,c,d) we assume $\mu=\mu_0$, $c=0.5$ and
$c^{\prime}=0.22$, and depict the plots for the LO and NLO QCD
corrected cross sections and the corresponding K-factors for the $pp
\to Z^0H^0+X$ process in both the SM and the LHM as the functions of
the global symmetry breaking scale $f$ at the $\sqrt{s}=14~TeV$ and
$\sqrt{s}=8~TeV$ LHC in Figs.\ref{fig7}(a) and (c), separately. The
corresponding relative deviations of the cross sections in the LHM
from those in the SM, $\delta(f)\equiv
\frac{\left[\sigma^{LHM}(f)-\sigma^{SM}(f)\right]}
{\sigma^{SM}(f)}$, are shown in Figs.\ref{fig7}(b) and (d),
respectively. We can see from Figs.\ref{fig7}(a,b,c,d) that when $f
\to \infty$, the relative deviations at both the LO and the NLO tend
to be vanished, and the relative deviations become to be less than
$5\%$ for the $\sqrt{s}=14~TeV$ LHC and the $\sqrt{s}=8~TeV$ LHC in
the ranges of $f>5~TeV$ and $f>4~TeV$, respectively. We find also
that the deviations are sensitive to the scale $f$ in the range of
$f< 5~TeV$ in both Figs.\ref{fig7}(b) and (d).

\par
\begin{figure}[htbp]
\includegraphics[scale=0.45]{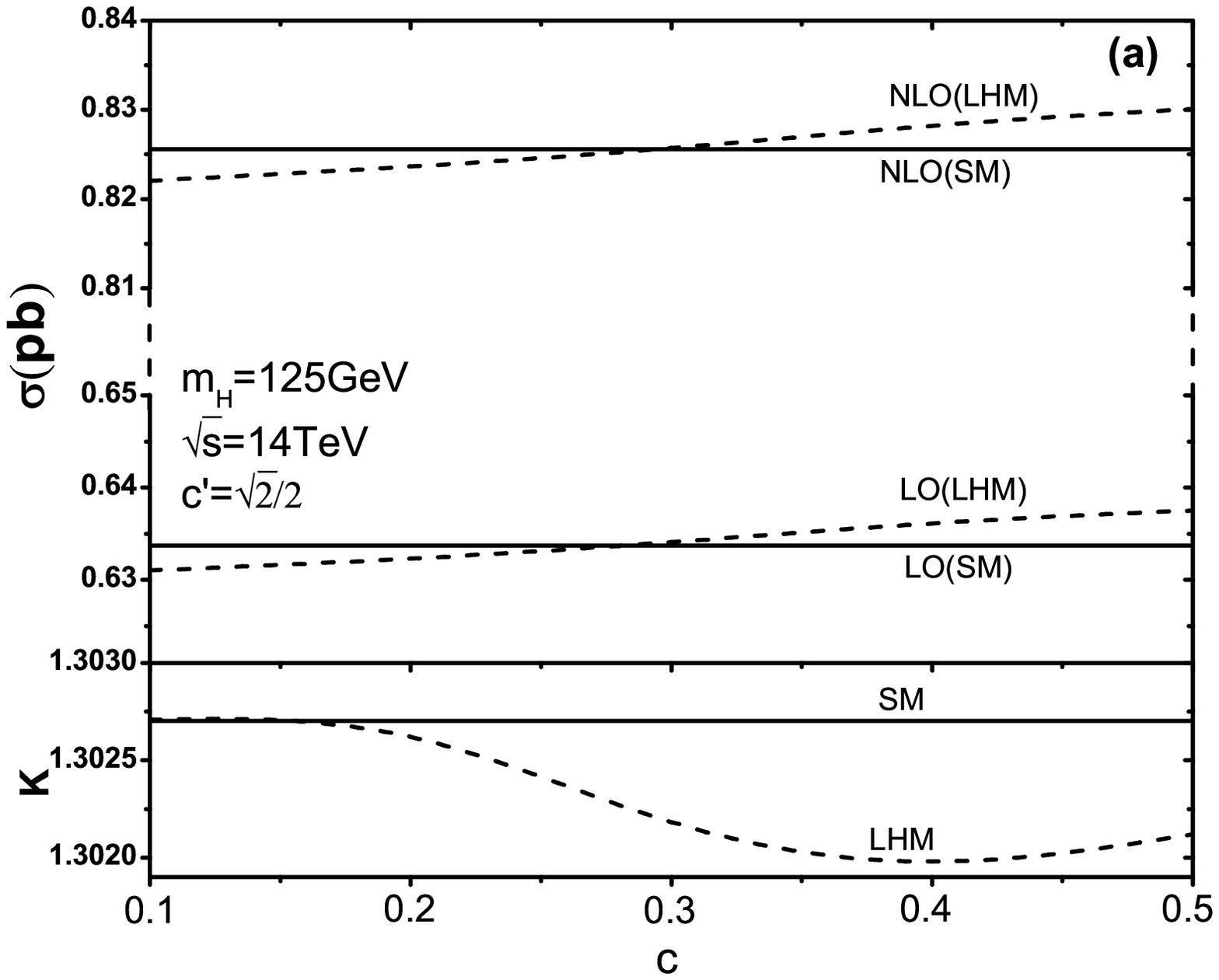}
\includegraphics[scale=0.45]{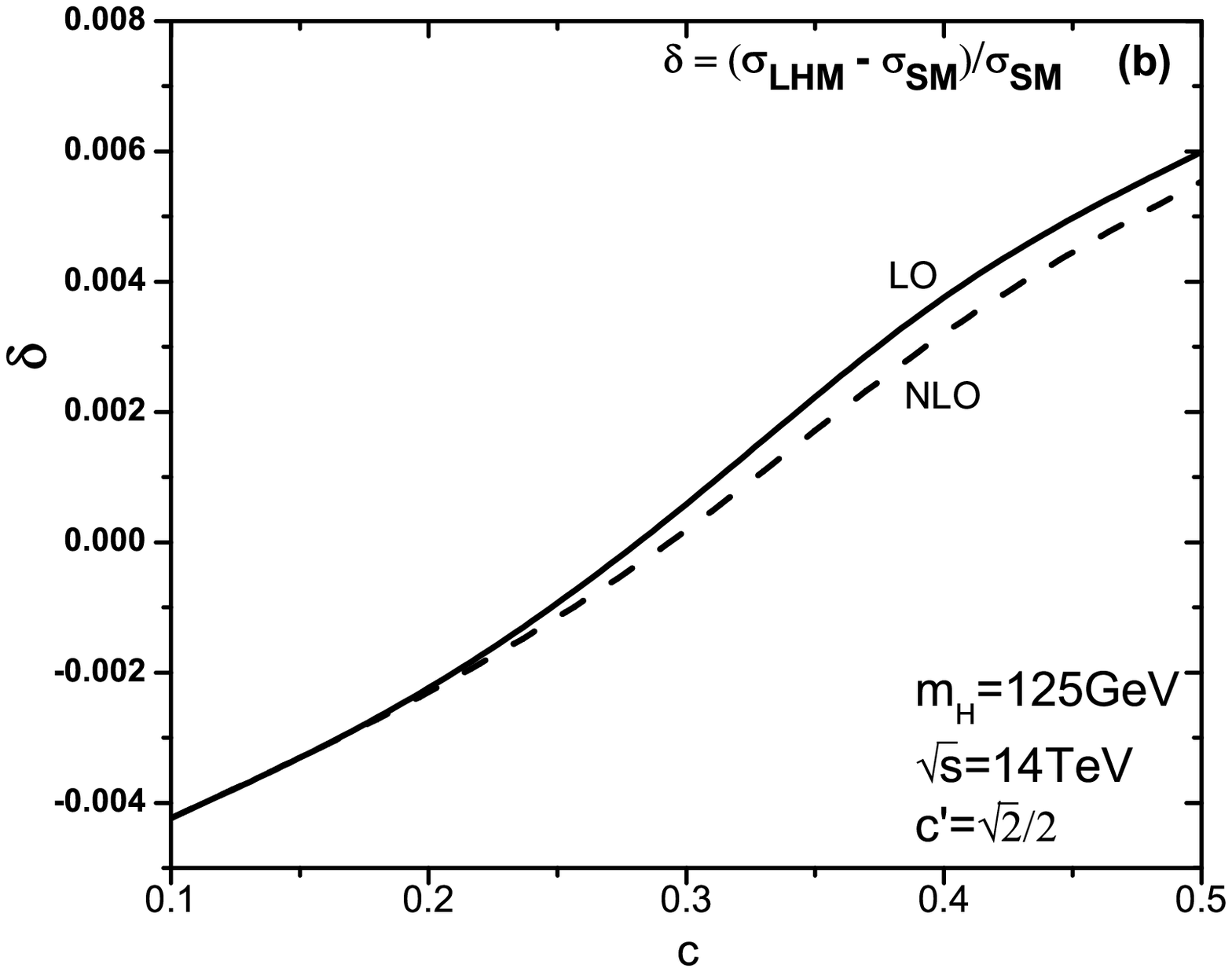}
\includegraphics[scale=0.45]{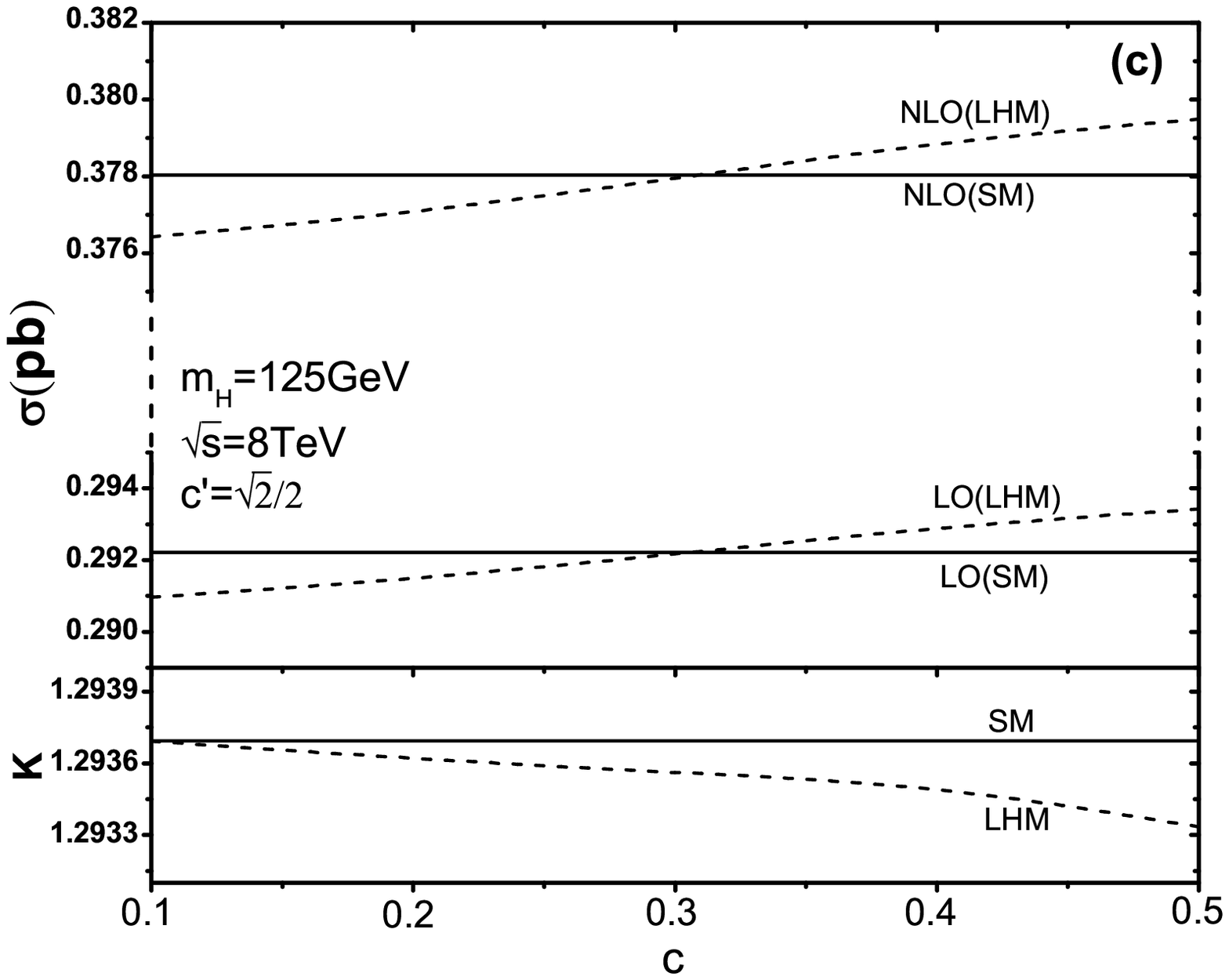}
\includegraphics[scale=0.45]{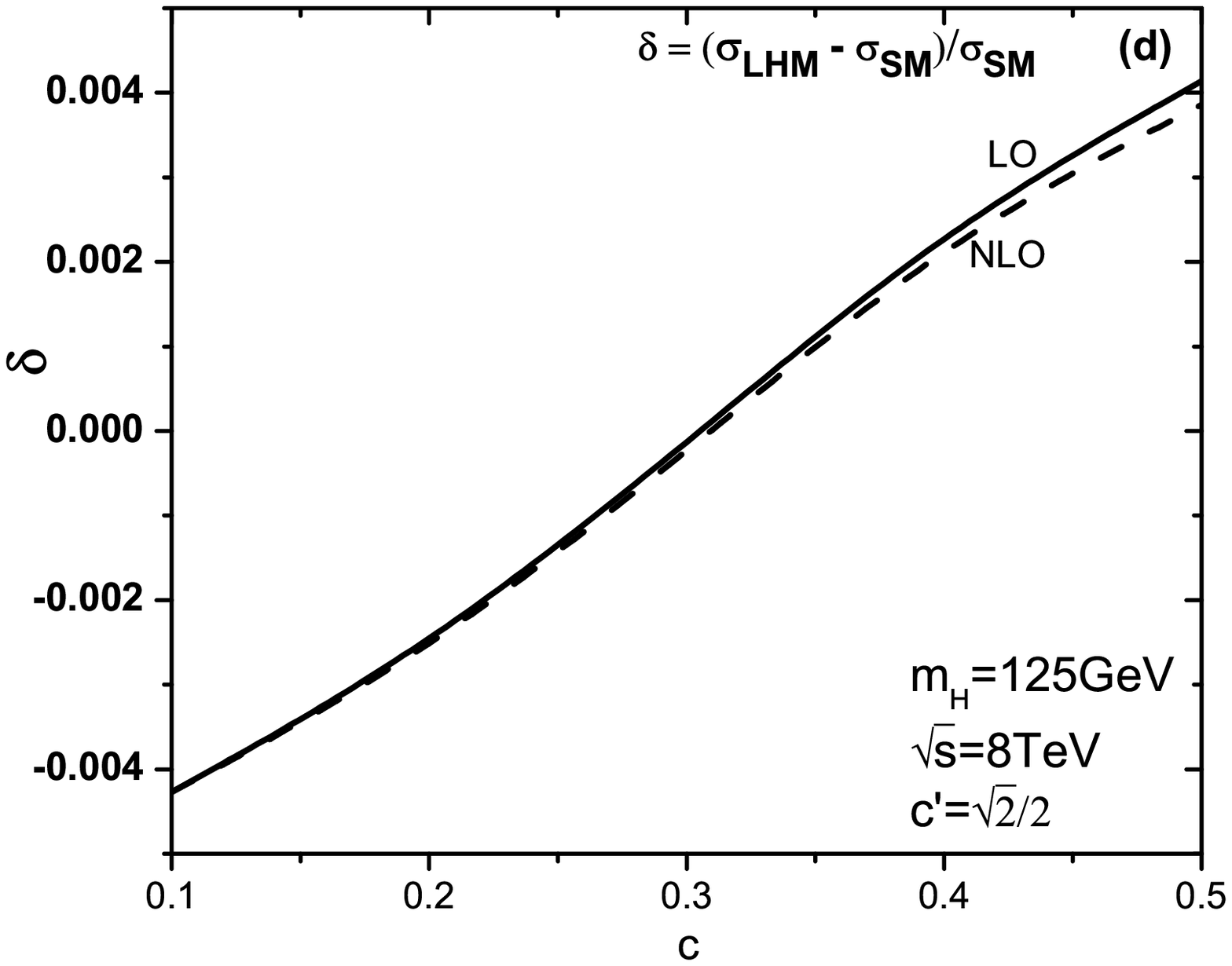}
\hspace{0in}%
\caption{\label{fig8} We take $c^{\prime}=1/\sqrt{2}$, $f=4~TeV$ and
$\mu=\mu_0$. (a) The LO and NLO QCD corrected cross sections and the
corresponding K-factors for the $pp \to Z^0H^0+X$ process at the
$\sqrt{s}=14~TeV$ LHC in both the SM and LHM as the functions of
parameter $c$. (b) The relative deviations of the cross sections in
the LHM from those in the SM corresponding to Fig.\ref{fig8} versus
parameter $c$. (c) The LO and NLO QCD corrected cross sections at
the $\sqrt{s}=8~TeV$ LHC in both the SM and LHM as the functions of
$c$. (d) The relative deviations corresponding to Fig.\ref{fig8}(c)
versus $c$. }
\end{figure}

From Eq.(\ref{ZH-mass}) we can conclude that the mass of the heavy
gauge boson $Z_H$ is mostly related with the scale $f$ and the
mixing angle parameter $c$ between two $SU(2)$ gauge bosons, but not
sensitive to the parameter $c^\prime$. For demonstrating the effects
from the interactions involving $Z_H$ boson, we take $\mu=\mu_0$,
$f=4~TeV$ and $c^{\prime}=1/\sqrt{2}$ in Figs.\ref{fig8}(a,b,c,d),
in which case the contributions from the $A_H$ exchange diagrams are
vanished (see Eqs.(\ref{Z_H-A_H-ZH})). We plot the LO and NLO QCD
corrected cross sections and the corresponding K-factors as the
functions of the parameter $c$ at the $\sqrt{s}=14~TeV$ and
$\sqrt{s}=8~TeV$ LHC in Figs.\ref{fig8}(a) and (c), separately. The
relative deviations of the cross sections in the LHM from those in
the SM, $\delta(c)\equiv
\frac{\left[\sigma^{LHM}(c)-\sigma^{SM}(c)\right]}
{\sigma^{SM}(c)}$, are shown in Figs.\ref{fig8}(b) and (d).
Figs.\ref{fig8}(a) and (c) show that the K-factors in the LHM and SM
are beyond $1.29$ for both the $\sqrt{s}=14~TeV$ and
$\sqrt{s}=8~TeV$ LHC. We can see from Figs.\ref{fig8}(b,d) that the
difference between the relative deviations of $\delta_{LO}(c)$ and
$\delta_{NLO}(c)$ goes up with the increment of the mixing angle
parameter $c$ in the range of $c\in [0.1,~0.5]$, and the LO and NLO
deviations in the LHM and SM are all sensitive to the mixing angle
parameter $c$. We see also that in the range of $c< 0.2$ the LO and
NLO relative deviations between the two models are nearly the same
for both the $\sqrt{s}=14~TeV$ and $\sqrt{s}=8~TeV$ LHC.

\par
\begin{figure}[htbp]
\includegraphics[scale=0.45]{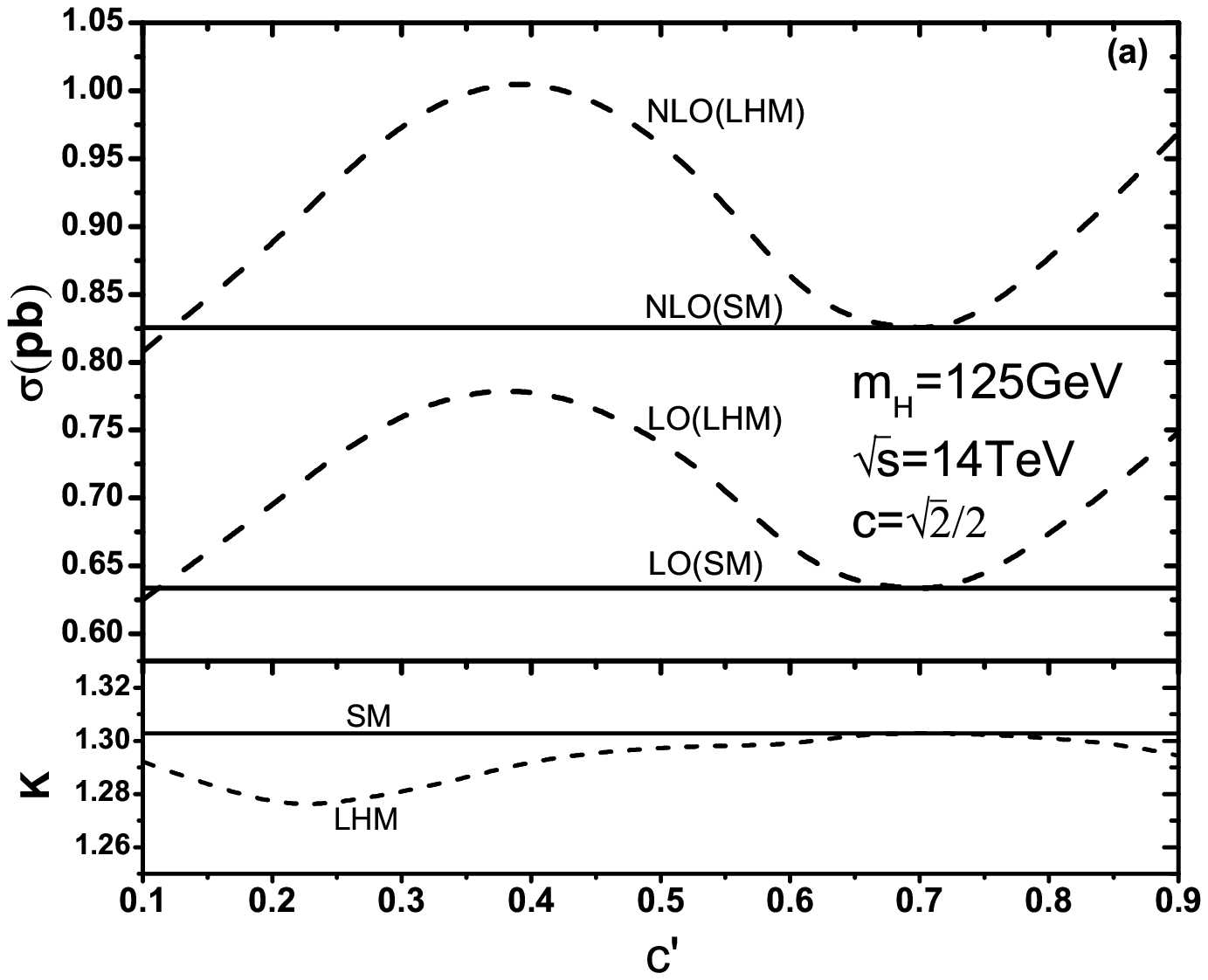}
\includegraphics[scale=0.45]{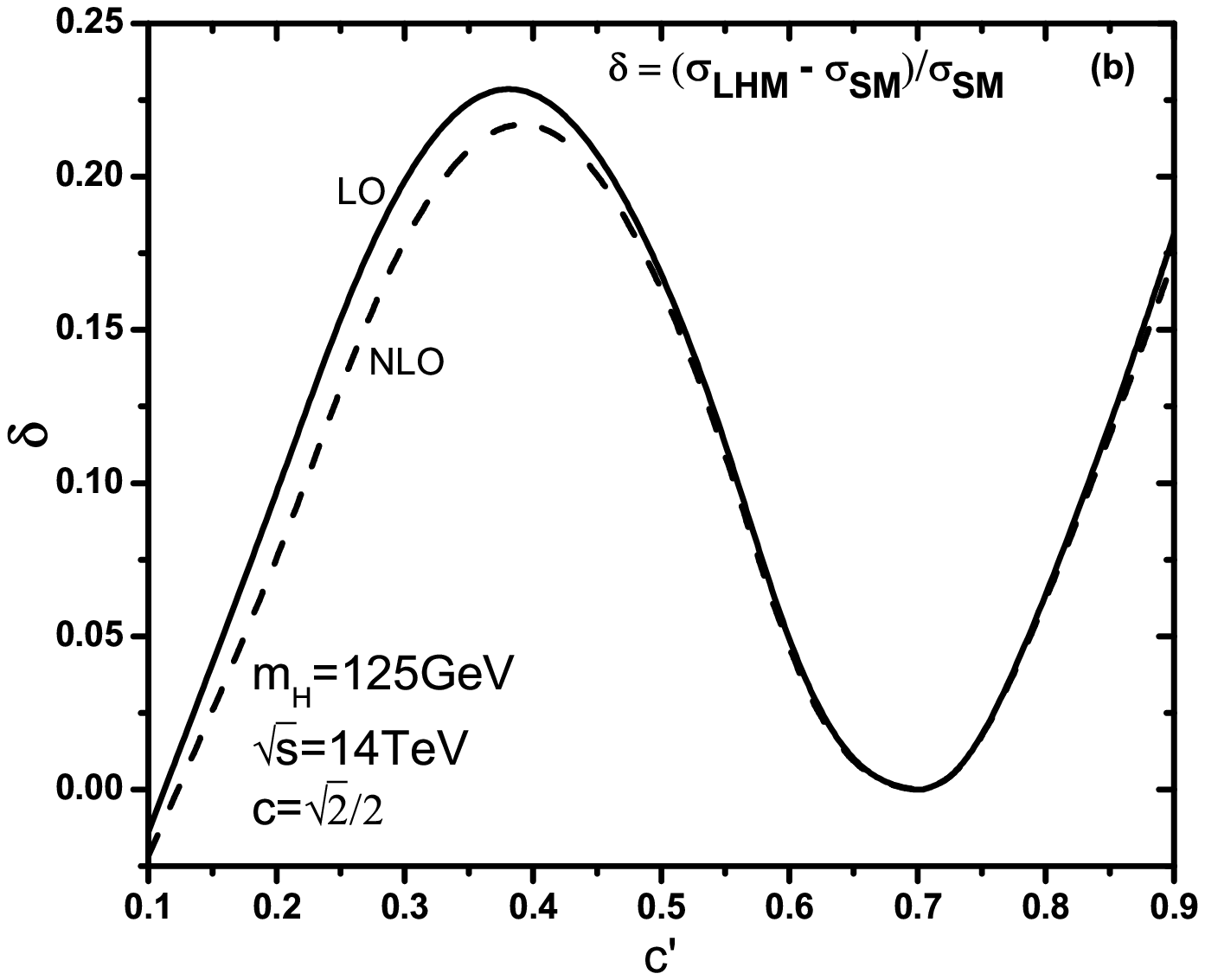}
\includegraphics[scale=0.45]{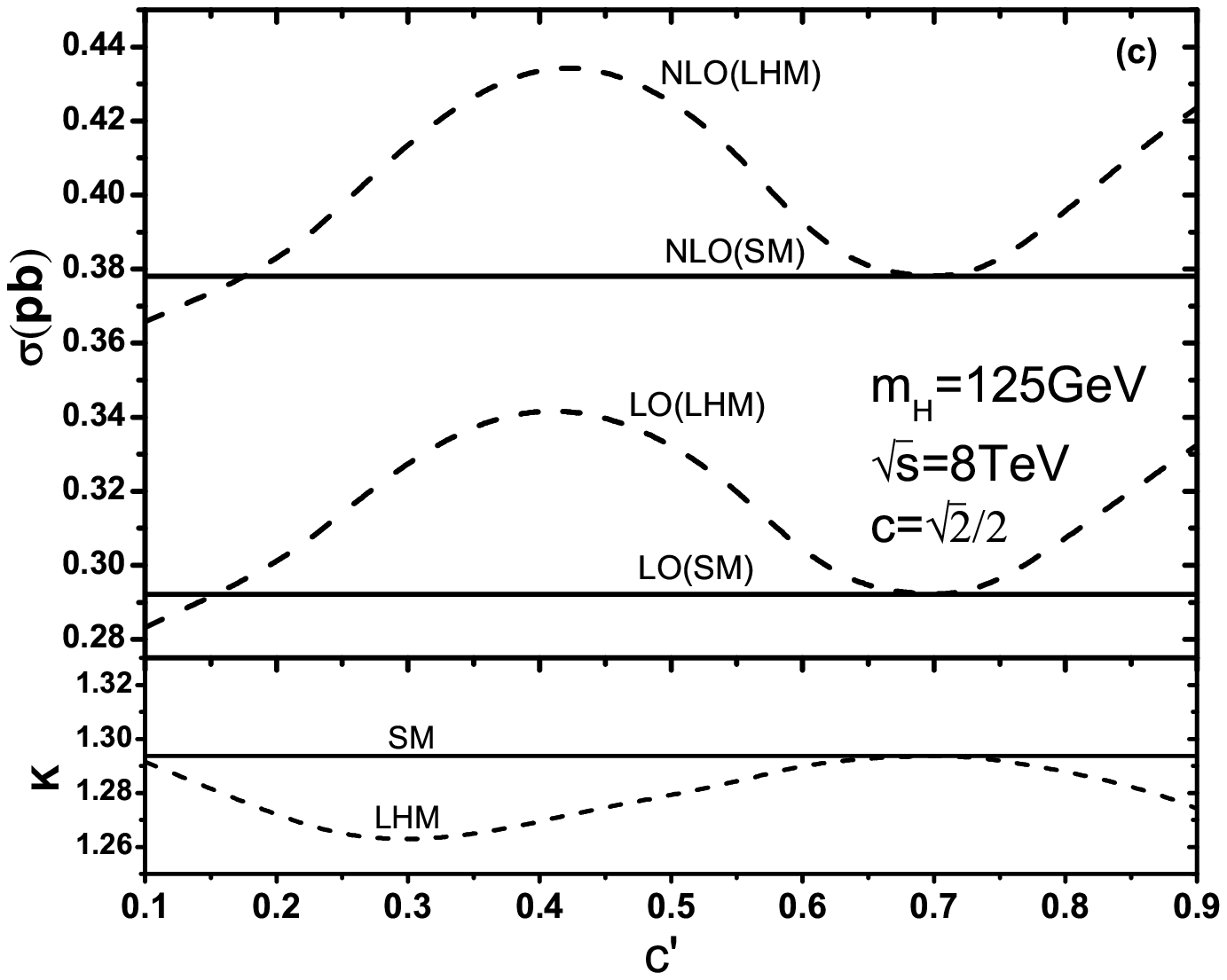}
\includegraphics[scale=0.45 ]{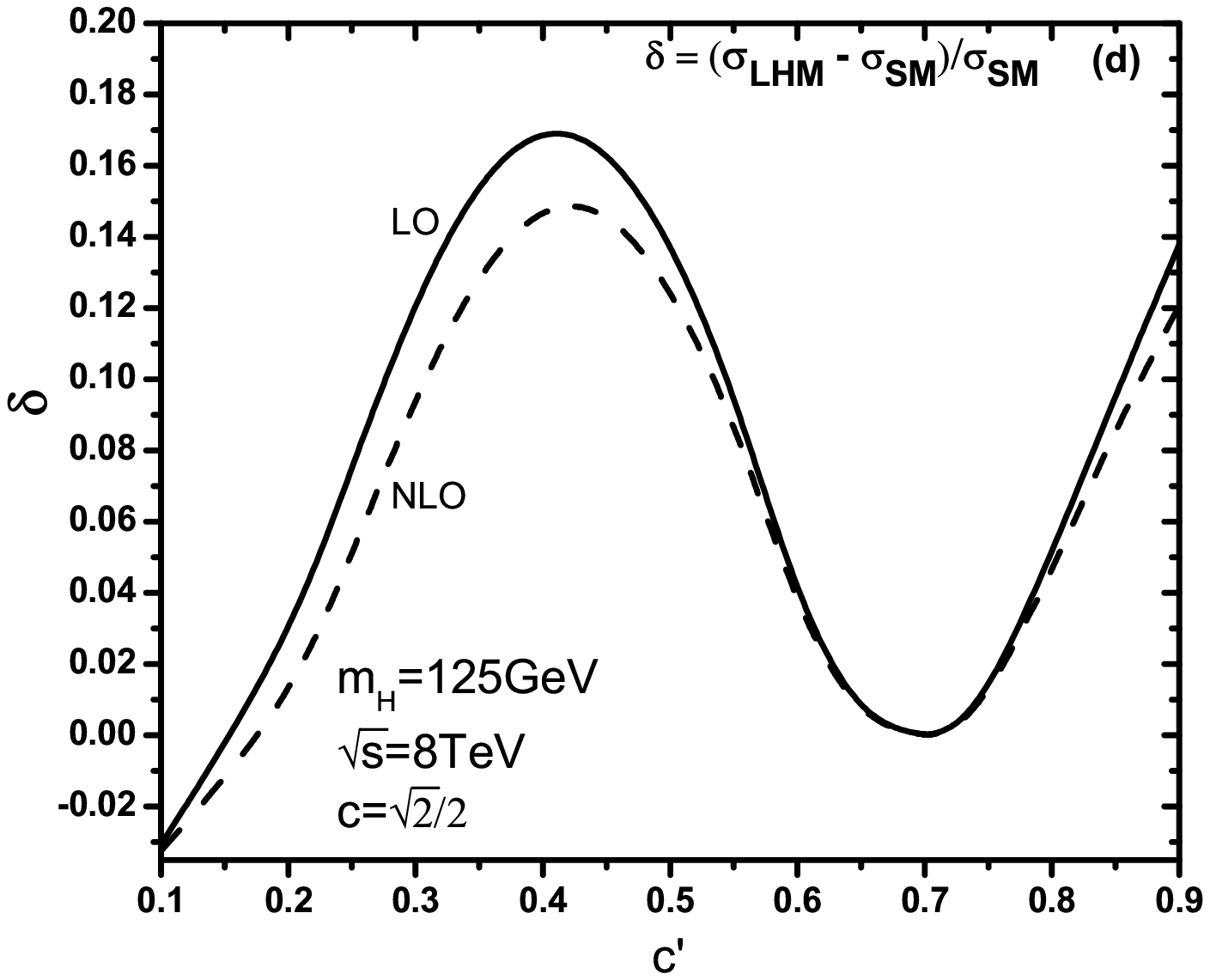}
\hspace{0in}%
\caption{\label{fig9} We take $c=1/\sqrt{2}$, $f=4~TeV$ and
$\mu=\mu_0$. (a) The LO and NLO QCD corrected cross sections and the
corresponding K-factors for the $pp \to Z^0H^0+X$ process at the
$\sqrt{s}=14~TeV$ LHC in both the SM and LHM as the functions of the
parameter $c^\prime$. (b) The relative deviations of the cross
sections in the LHM from those in the SM corresponding to
Fig.\ref{fig9}(a) versus parameter $c^\prime$. (c) The LO and NLO
QCD corrected cross sections at the $\sqrt{s}=8~TeV$ LHC in both the
SM and LHM as the functions of $c^\prime$. (d) The relative
deviations corresponding to Fig.\ref{fig9}(c) versus parameter
$c^\prime$. }
\end{figure}

Eq.(\ref{AH-mass}) tells us that the heavy photon mass $M_{A_H}$
mainly depends on the scale $f$ and the mixing angle parameter
$c^\prime$ between two $U(1)$ gauge fields, but is insensitive to
the mixing parameter $c$. In order to investigate and discuss the
contributions of $A_H$ exchange diagrams to the $Z^0H^0$ associated
production, we present the LO and NLO QCD corrected cross sections
and the corresponding K-factors as the functions of the mixing angle
parameter $c^\prime$ at the $\sqrt{s}=14~TeV$ and $\sqrt{s}=8~TeV$
LHC in Figs.\ref{fig9}(a) and (c), separately. In
Figs.\ref{fig9}(a,b,c,d) we take $\mu=\mu_0$, $f=4~TeV$ and
$c=1/\sqrt{2}$, in this case there is no contribution from the $Z_H$
exchange diagrams (see Eq.(\ref{Z_H-A_H-ZH})). The corresponding
relative deviations of the cross sections in the LHM from those in
the SM, $\delta(c^\prime)\equiv \frac{\left[\sigma^{LHM}(c^\prime)-
\sigma^{SM}(c^\prime)\right]} {\sigma^{SM}(c^\prime)}$, are
demonstrated in Figs.\ref{fig9}(b) and (d), respectively. We can see
from Figs.\ref{fig9}(a) and (c) that the LO and NLO QCD corrected
total cross sections in the LHM at the early and future LHC are
obviously related to the mixing angle parameter $c^\prime$ in the
range of $c^\prime \in [0.10,~0.65]$, and the K-factors in the LHM
are sensitive to $c^\prime$ in the range of $c^\prime \in
[0.10,~0.65]$. Figs.\ref{fig9}(b) and (d) demonstrate that the
difference of $\delta_{LO}(c^\prime)-\delta_{NLO}(c^\prime)$ becomes
smaller when $c^\prime$ increases from $0.40$ to $0.65$, while in
the range of $c^\prime \in [0.65,~0.90]$ the NLO and LO relative
deviations, $\delta_{NLO}(c^\prime)$ and $\delta_{LO}(c^\prime)$,
have almost the same values for the $\sqrt{s}=14~TeV$ and
$\sqrt{s}=8~TeV$ LHC.

\par
\begin{figure}[htbp]
\includegraphics[scale=0.55]{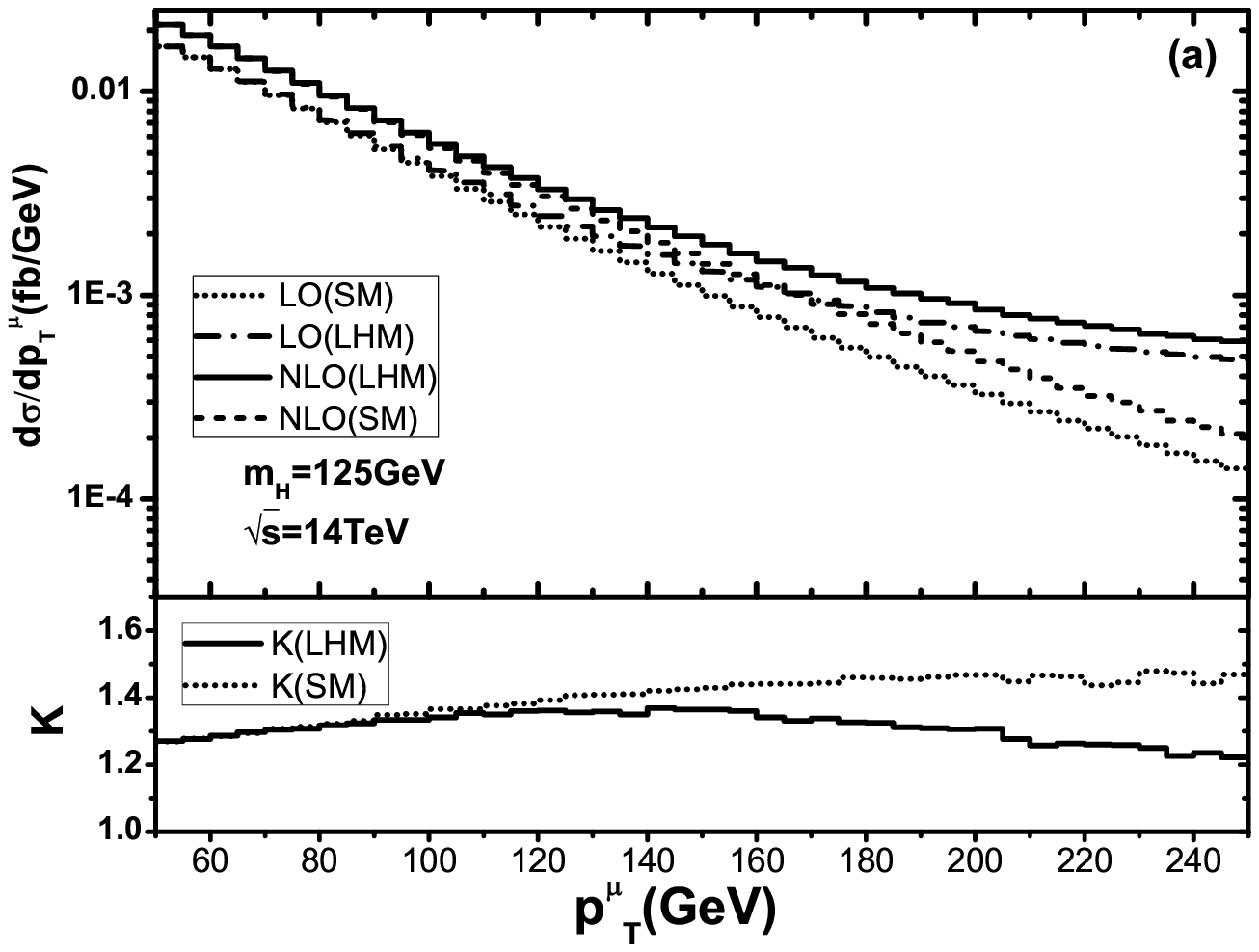}
\includegraphics[scale=0.55]{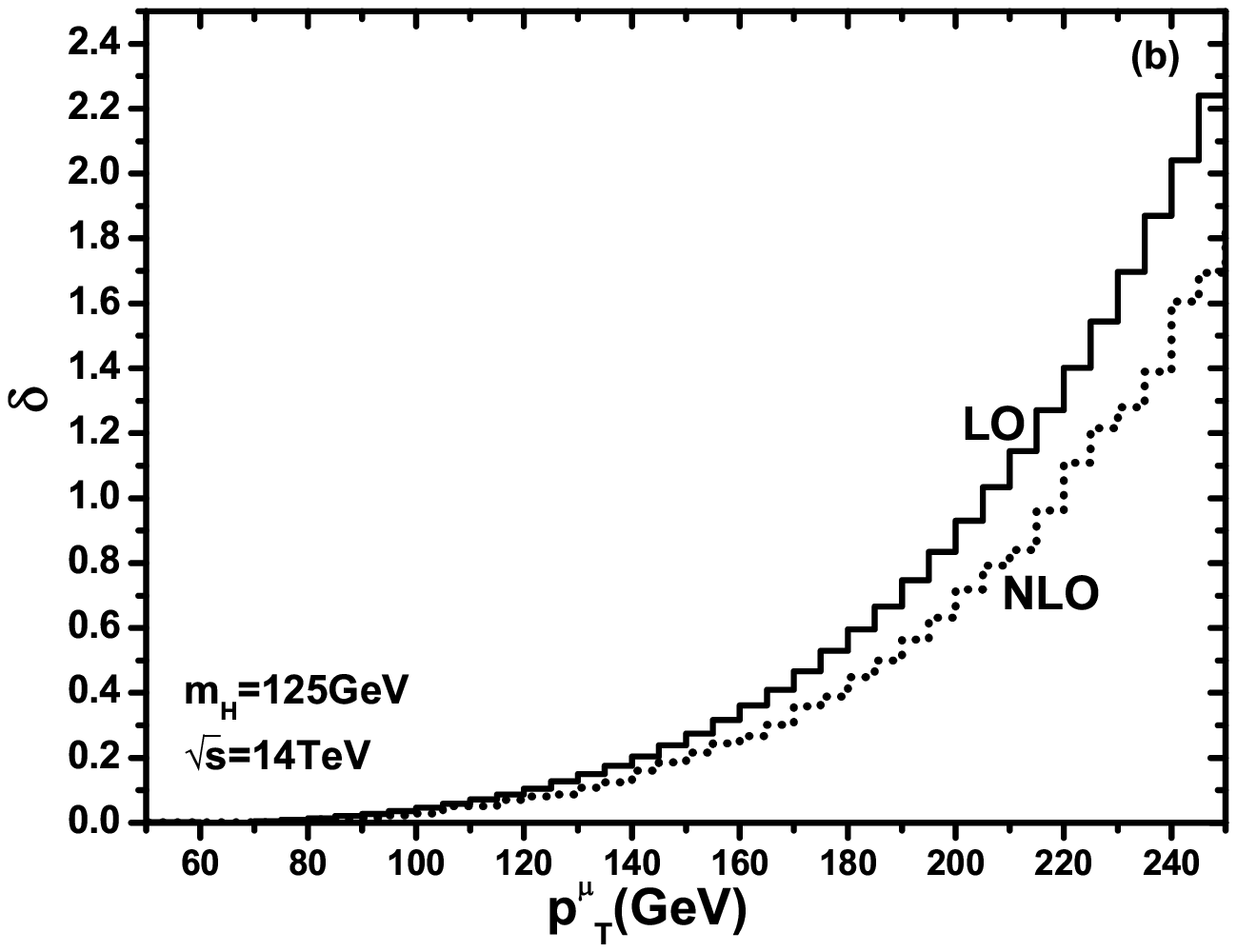}
\includegraphics[scale=0.55]{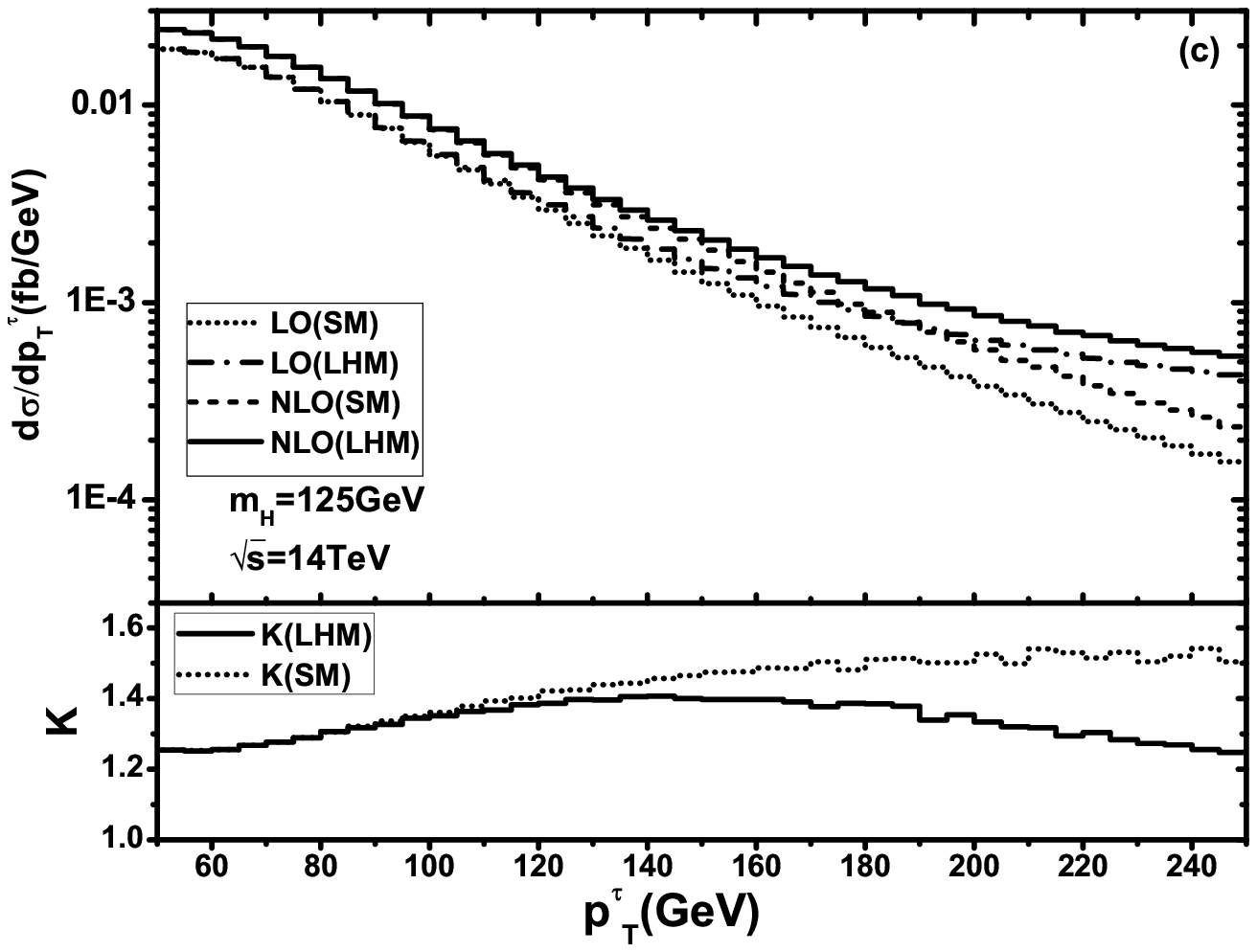}
\includegraphics[scale=0.55]{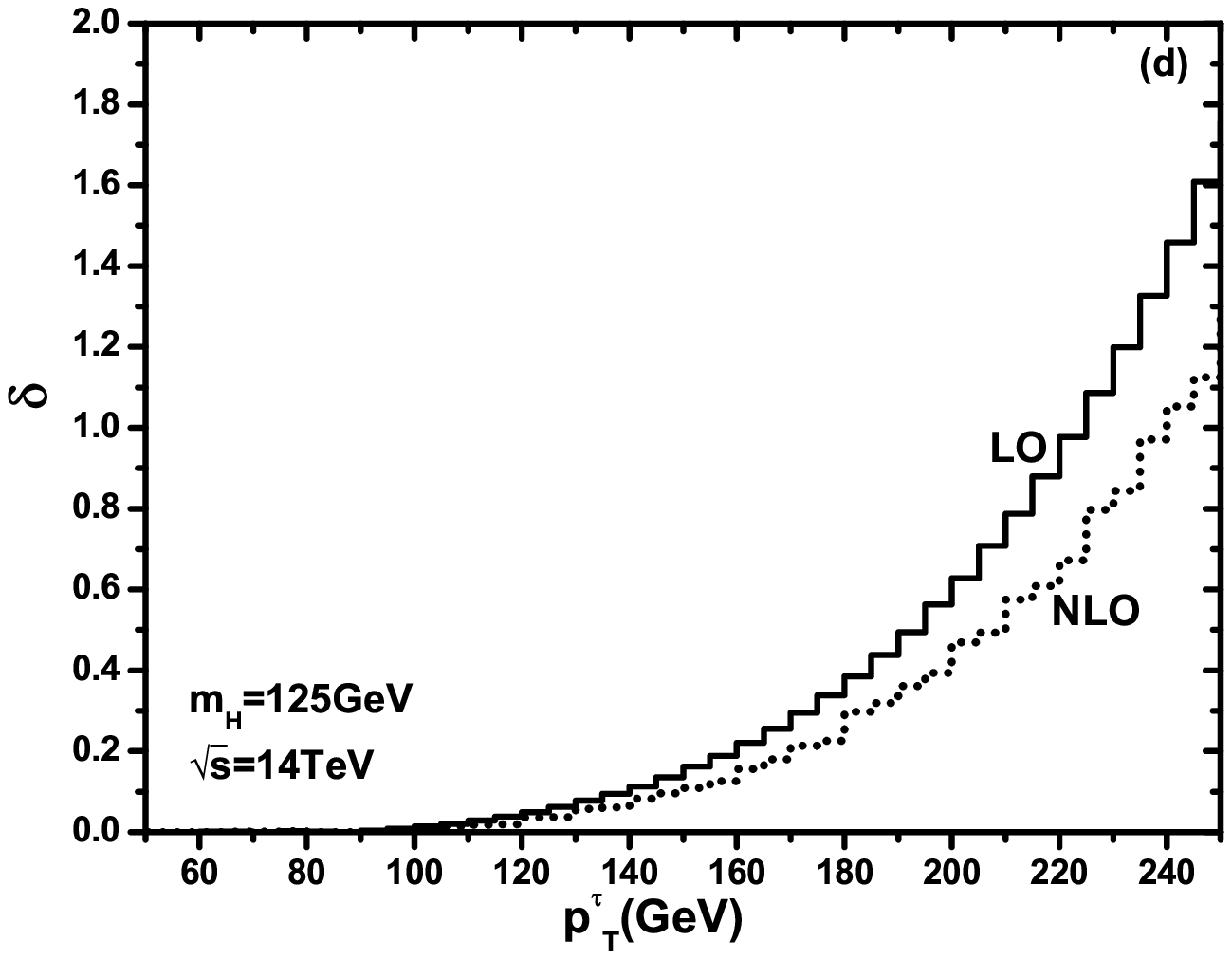}
\hspace{0in}%
\caption{\label{fig10} The LO and NLO QCD corrected distributions of
the transverse momenta of final leptons and the corresponding
K-factors for the $pp \to Z^0H^0 \to \mu^+\mu^-\tau^+\tau^-+X$
process at the $\sqrt{s}=14~TeV$ LHC in both the SM and LHM, where
we take $\mu=\mu_0$, $c=0.5$, $c^{\prime}=0.22$ and $f=4~TeV$. (a)
The distributions of $p_T^{\mu^-}$. (b) The relative deviations of
the cross sections in the LHM from those in the SM corresponding to
Fig.\ref{fig10}(a) as the functions of $p_T^{\mu^-}$. (c) The
distributions of $p_T^{\tau^-}$. (d) The relative deviations
corresponding to Fig.\ref{fig10}(c) as the functions of
$p_T^{\tau^-}$. }
\end{figure}

As we know, the final $Z^0$ and $H^0$ bosons are unstable and can be
detected experimentally via the subsequential leptonic decays of
$Z^0 \to \mu^+\mu^-$ and $H^0 \to \tau^+\tau^-$. We employ the SM
leptonic decay branch ratios of $Z^0$ and $H^0$ boson in further
numerical calculations, i.e., $Br(Z^0 \to \mu^+\mu^-)=3.366\%$ and
$Br(H^0 \to \tau^+\tau^-)=6.5\%$ \cite{hdata}. Since the transverse
momentum distributions of $\mu^+$ and $\tau^+$ of the process $pp
\to Z^0H^0 \to \mu^+\mu^-\tau^+\tau^-+X$ should be the same as those
of $\mu^-$ and $\tau^-$ correspondingly, we present only those of
$\mu^-$ and $\tau^-$. We depict the LO and QCD NLO corrected
transverse momentum distributions of final $\mu^-$ and the
corresponding relative deviations of the cross sections in the LHM
from those in the SM at the $\sqrt{s}=14~TeV$ LHC in
Figs.\ref{fig10}(a) and (b) separately, where we take $M_H=125~GeV$,
$c=0.5$, $c^{\prime}=0.22$ and $f=4~TeV$.  All the curves in
Figs.\ref{fig10}(a) go down with the increment of the
$\mu^-$ transverse momentum within the plotted $p_T^{\mu^-}$ range.
The differential cross sections, $d\sigma_{LO,NLO}/dp^{\tau^-}_T$,
and the corresponding relative
deviations of the cross sections at the $\sqrt{s}=14~TeV$ LHC in the LHM
from those in the SM as the functions of $p_T^{\tau^-}$ are shown in
Figs.\ref{fig10}(c) and (d), respectively. There we adopt again $M_H=125~GeV$,
$c=0.5$, $c^{\prime}=0.22$ and $f=4~TeV$. In Fig.\ref{fig10}(c) we
see that the curves for the LO and QCD NLO distributions of $p_T^{\tau^-}$
in both the SM and the LHM frameworks fall down
when the transverse momentum $p_T^{\tau^-}$ goes up.
Figs.\ref{fig10}(a,c) demonstrate that the LO differential cross
sections of $d\sigma_{LO}/dp^{\mu^-}_T$ and $d\sigma_{LO}/dp^{\tau^-}_T$
in both the SM and the LHM frameworks are
significantly enhanced by the QCD NLO corrections. Fig.\ref{fig10}(b)
and Fig.\ref{fig10}(d) show that the corresponding relative
deviations between the two models are significantly suppressed by
the QCD NLO corrections, and in the ranges of $p_T^{\mu^-} > 130~GeV$
and $p_T^{\tau^-} > 150~GeV$ the QCD NLO corrected relative deviations
can exceed $10\%$, separately.

\par
\begin{figure}[htbp]
\includegraphics[scale=0.55]{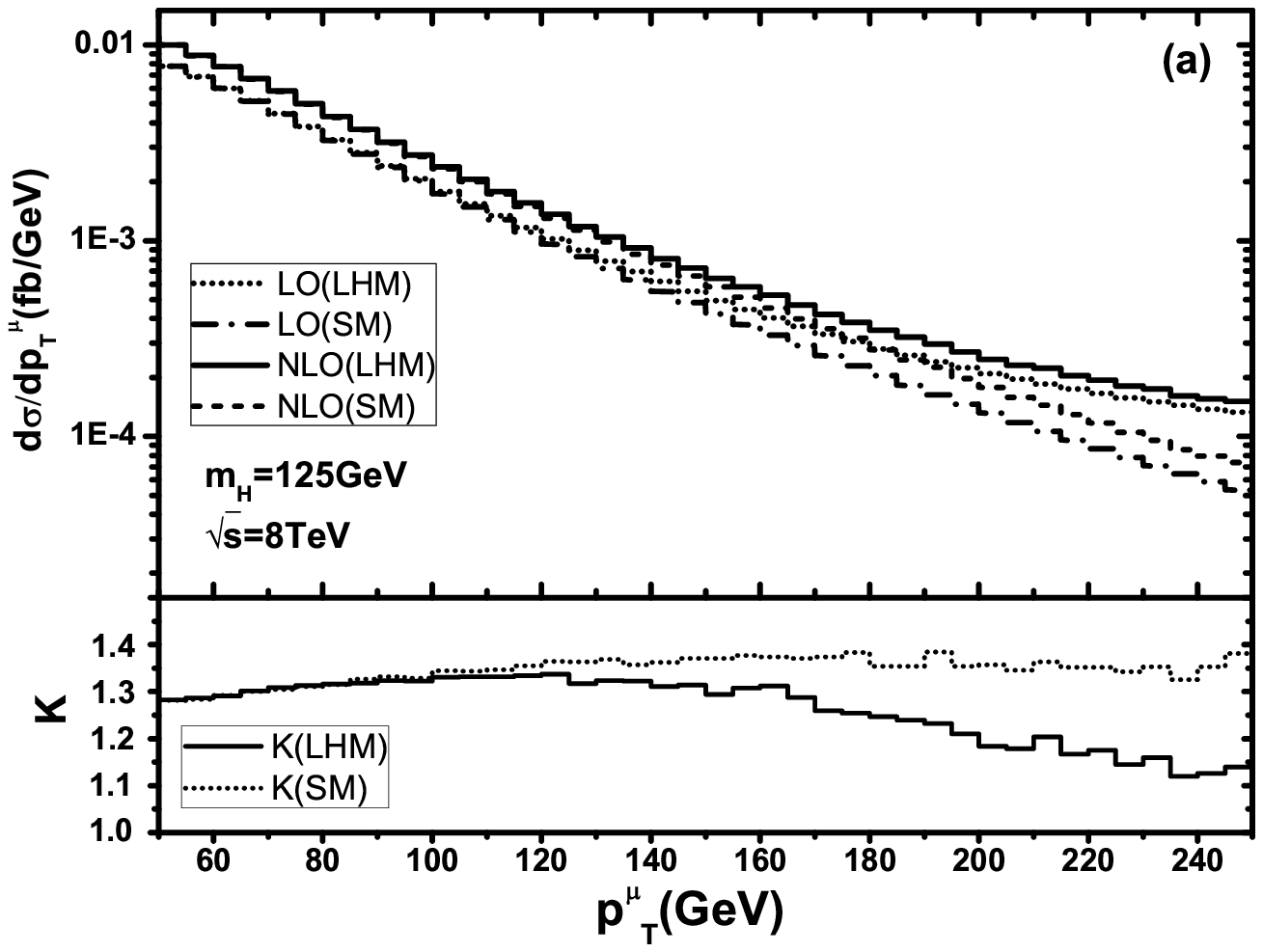}
\includegraphics[scale=0.55]{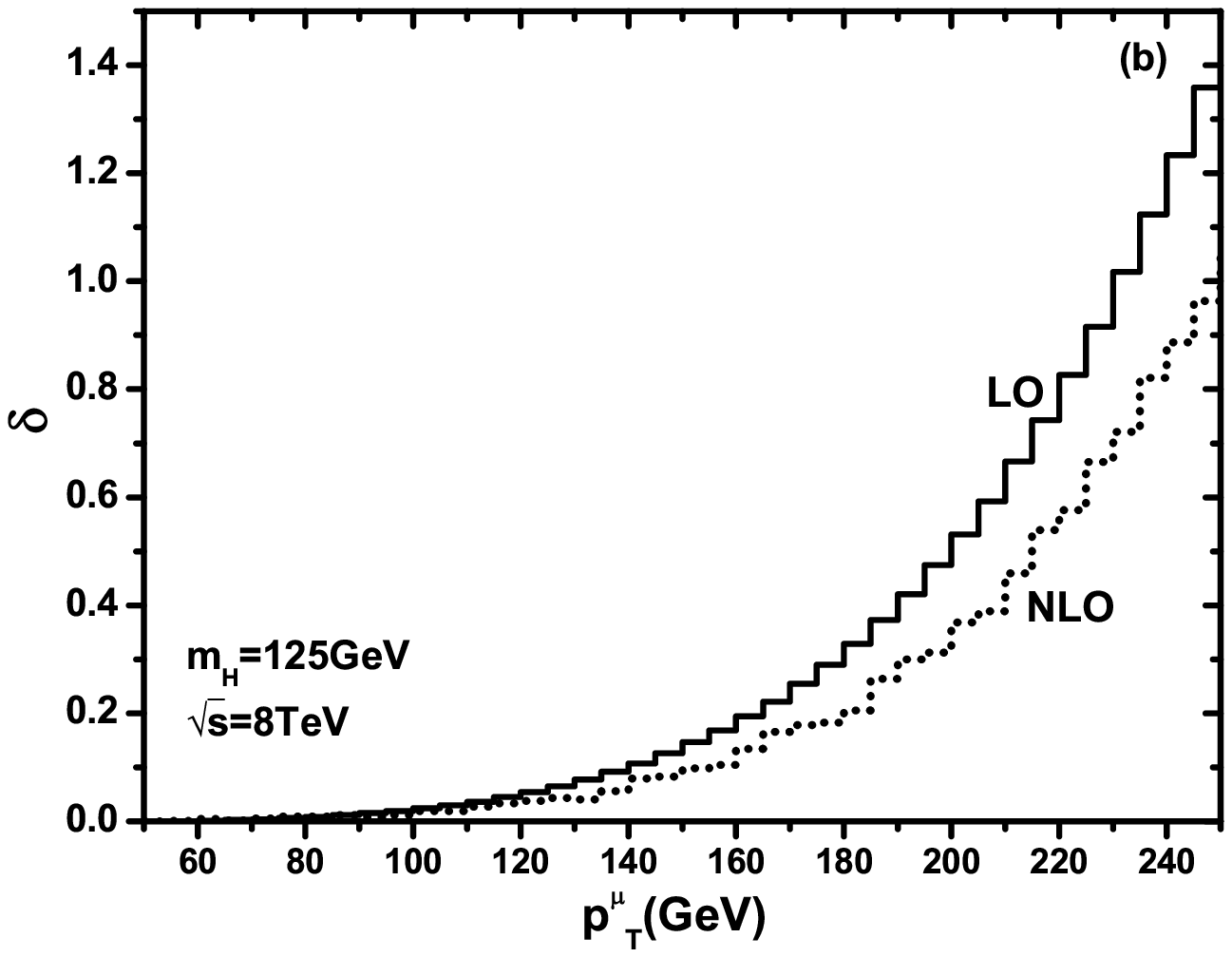}
\includegraphics[scale=0.55]{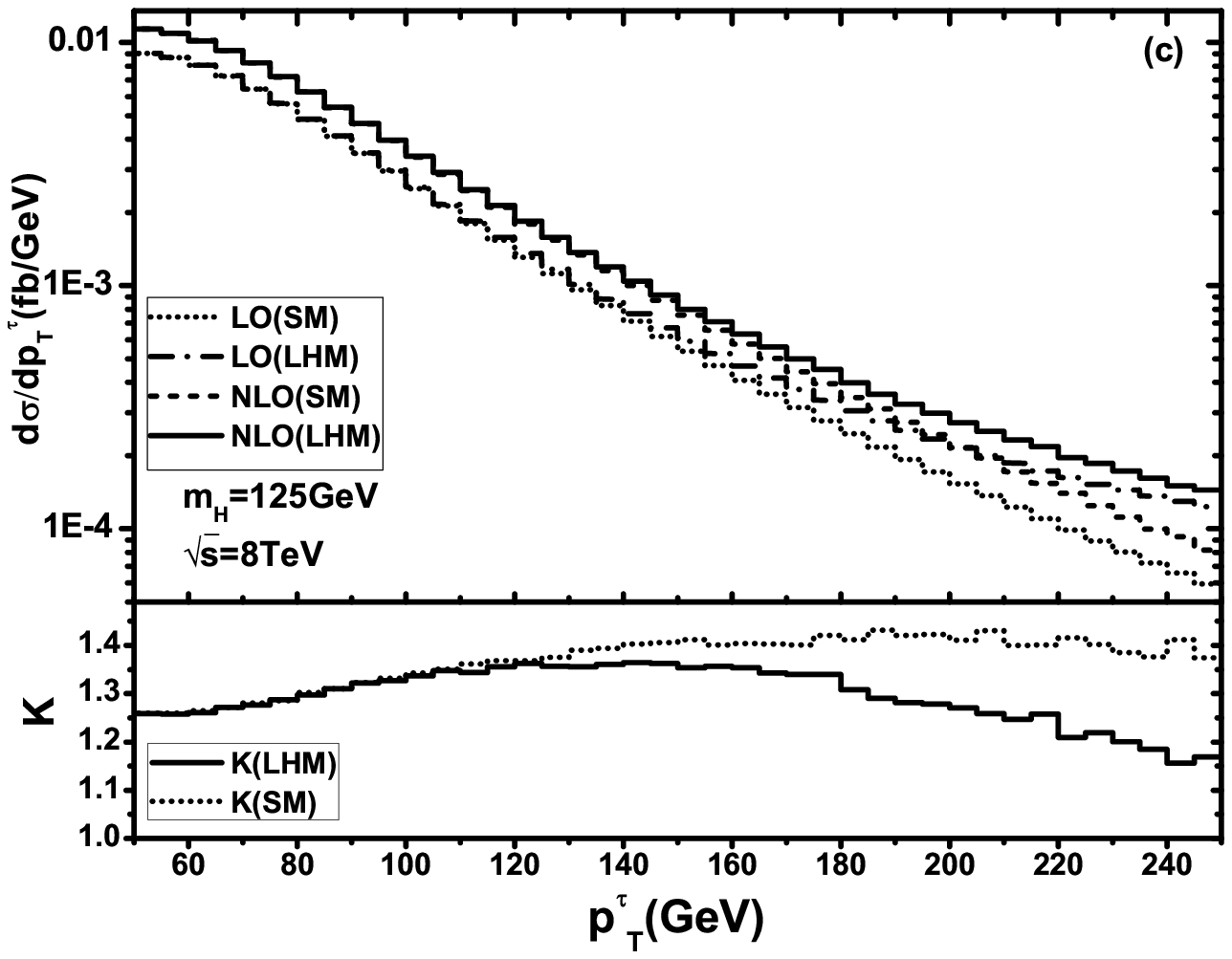}
\includegraphics[scale=0.55]{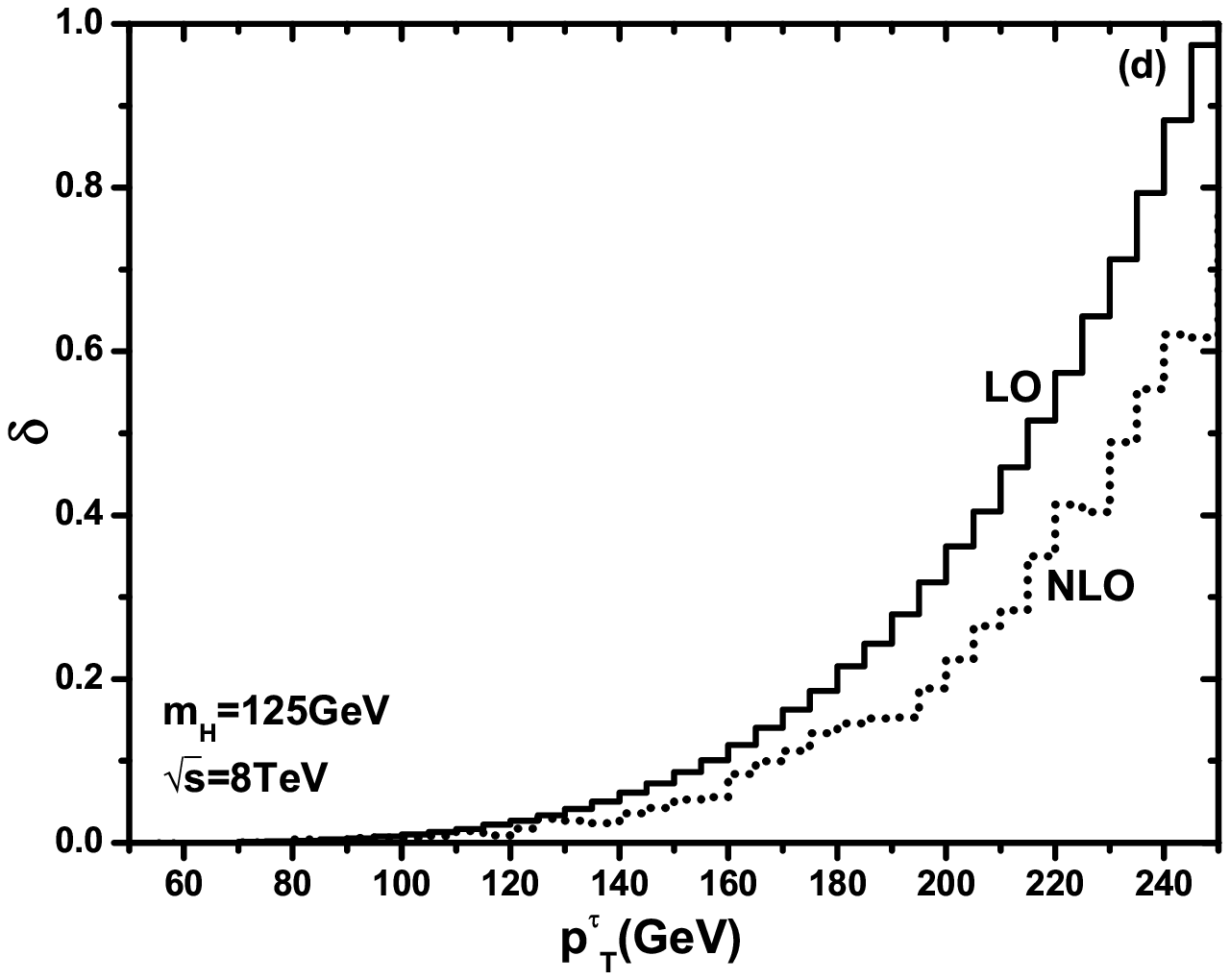}
\hspace{0in}%
\caption{\label{fig11}  The LO and NLO QCD corrected distributions
of the transverse momenta of final leptons and the corresponding
K-factors for the $pp \to Z^0H^0 \to \mu^+\mu^-\tau^+\tau^-+X$
process at the $\sqrt{s}=8~TeV$ LHC in both the SM and LHM, where we
take $c=0.5$, $c^{\prime}=0.22$ and $f=4~TeV$. (a) The distributions
of $p_T^{\mu^-}$. (b) The relative deviations of the differential
cross sections in the LHM from those in the SM corresponding to
Fig.\ref{fig11}(a) as the functions of $p_T^{\mu^-}$. (c) The
distributions of $p_T^{\tau^-}$. (d) The relative deviations
corresponding to Fig.\ref{fig11}(c) as the functions of
$p_T^{\tau^-}$. }
\end{figure}

Similar with Figs.\ref{fig10}(a,b,c,d) we plot the corresponding
distributions of final $\mu^-$ and $\tau^-$ in the $pp \to Z^0H^0
\to \mu^+\mu^-\tau^+\tau^-+X$ process at the $\sqrt{s}=8~TeV$ LHC in
Figs.\ref{fig11}(a,b,c,d). From Fig.\ref{fig11}(a) and
Fig.\ref{fig11}(c) we can see that for the $\sqrt{s}=8~TeV$ LHC
all the curves for both the LO and QCD NLO distributions of
$p_T^{\mu^-}$ and $p_T^{\tau^-}$ decrease with the increment of the
corresponding transverse momentum, which are similar with the curves for the
$\sqrt{s}=14~TeV$ LHC. Again, we see that both the LO differential
cross sections of $p^{\mu^-}_T$ and $p^{\tau^-}_T$
($d\sigma_{LO}/dp^{\mu^-}_T$, $d\sigma_{LO}/dp^{\tau^-}_T$) are
significantly enhanced by the QCD corrections. We can see from
Fig.\ref{fig11}(b) and Fig.\ref{fig11}(d) that the relative
deviations between the two models are significantly suppressed by
the QCD NLO corrections, and in the ranges of $p_T^{\mu^-} > 155~GeV$
and $p_T^{\tau^-}> 170~GeV$ the QCD NLO corrected deviations at the
$\sqrt{s}=8~TeV$ LHC can exceed $10\%$, separately.

\vskip 5mm
\par
\section{Summary}
\par
In this paper we investigate the phenomenological effects induced by
the new heavy neutral gauge bosons in the LHM up to QCD NLO on the
$Z^0H^0$ associated production at the early ($\sqrt{s}=8~TeV$) and
future ($\sqrt{s}=14~TeV$) LHC. We study the dependences of the LO
and NLO QCD corrected cross sections on the
factorization/renormalization scale $\mu$, the LHM parameters $c$,
$c'$ and $f$, and present the LO and NLO QCD corrected distributions
of the transverse momenta $p_T^{\mu^-}$ and $p_T^{\tau^-}$. It
demonstrates that the new neutral gauge bosons could induce
significant discrepancies to the kinematic observables from the
standard model predictions for this process at both LO and up to QCD
NLO. Our results show that when we take the $c=0.5$,
$c^{\prime}=0.22$, $f=4~TeV$ and $\mu=\mu_0$, the effects from the
heavy neutral gauge boson interactions can make the relative
deviations to be about $12.83\%$ and $10.37\%$ at the LO and up to
QCD NLO, respectively. We find that the QCD corrections at the
$\sqrt{s}=14~TeV$ LHC can obviously make the cross section being
mildly related to the $\mu$ scale, and significantly enhance the
differential cross sections of the transverse momenta of the final
decay products $\mu$ and $\tau$. We also find the LO relative
deviations of the integrated cross sections are significantly
suppressed by the NLO QCD corrections. We conclude that the
precision measurement of the $Z^0H^0$ associated production process
at the LHC could provide the clue of the LHM physics.

\vskip 5mm
\par
\noindent{\large\bf Acknowledgments:} This work was supported in
part by the National Natural Science Foundation of China (Contract
No.11075150, No.11005101), and the Specialized Research Fund for the
Doctoral Program of Higher Education (Contract No.20093402110030).


\vskip 5mm
\begin{thebibliography}{99}
\bibitem{sm1}
   S. L. Glashow, Nucl. Phys. {\bf 22}, (1961) 579; S. Weinberg, Phys.
   Rev. Lett. {\bf 19}, (1967) 1264; A. Salam, Proc. 8th Nobel Symposium
   Stockholm 1968, ed. N. Svartholm (Almquist and Wiksells, Stockholm
   1968) p.367; H. D. Politzer, Phys. Rep. {\bf 14}, (1974) 129.

\bibitem{sm2}
   P. W. Higgs, Phys. Lett. {\bf 12}, (1964) 132; Phys. Rev. Lett. {\bf
   13}, (1964) 508; Phys. Rev. {\bf 145}, 1156 (1966); F. Englert and
   R. Brout, Phys. Rev. Lett. {\bf 13}, 321 (1964); G. S. Guralnik, C.
   R. Hagen and T. W. B. Kibble, Phys. Rev. Lett. {\bf 13}, (1964) 585;
   T. W. B. Kibble, Phys. Rev. {\bf 155}, (1967) 1554.

\bibitem{mh-Fermilab}
    The CDF, D0 Collaborations, the Tevatron New Phenomena, Higgs Working Group,
    {\it Combined CDF and D0 Upper Limits on Standard Model Higgs Boson Production
    with up to $8.6~fb^{-1}$ of Data}, FERMILAB-CONF-11-354-E, arXiv:1107.5518.

\bibitem{mh-Atlas}
    ATLAS Collaboration, [https://twiki.cern.ch/twiki/bin/view/AtlasPublic/AtlasResultsEPS2011].

\bibitem{mh-CMS}
    CMS Collaboration,{\it Combination of Higgs Searches}, CMS PAS HIG-11-022,
    [http://cms.web.cern.ch/cms/News/2011/LP11].

\bibitem{mh-Atlas-1}
    ATLAS Collaboration, 'ATLAS experiment presents latest Higgs search status',
    [http://www.atlas.ch/news/2011/status-report-dec-2011.html].

\bibitem{mh-CMS-1}
    CMS Collaboration,``CMS search for the Standard Model Higgs Boson in LHC
    data from 2010 and 2011'',
    [http://cms.web.cern.ch/news/cms-search-standard-model-higgs-boson-lhc-data-2010-and-2011].

\bibitem{hie}
N. Arkani-Hamed, A.G. Cohen and H. Georgi, Phys. Lett. {\bf B513},
232(2001); Phys. Lett. {\bf B513}, 232(2001); N. Arkani-Hamed, A.G.
Cohen, E.Katz, A.E. Nelson, T. Gregoire and J.G. Wacker, JHEP{\bf
0208} (2002) 021; M. Perelstain, Prog. Part. Nucl. Phys. 58 (2007)
247, arXiv:hep-ph/0512128.

\bibitem{super}
S. Dimopoulos and H. Georgi, Nucl. Phys. {\bf B193} (1981) 150; H.
P. Nilles, Phys. Rept. {\bf110} (1984) 1; H. E. Haber and G. L.
Kane, Phys. Rept. {\bf 117}(1985)75; S. P. Martin,
arXiv:hep-ph/9709356; P. Fayet. Nucl. Phys. {\bf B101} (Proc.
Suppl.) (2001) 81.

\bibitem{extra}
I. Antoniadis, C. Munoz, M. Quiros, Nucl. Phys. {\bf B397} (1993)
515; N. Arkani-Hamed, S. Dimopoulos, G. R. Dvali, Phys. Rev. {\bf
D59}1999) 086004; L. Randall, R. Sundrum, Phys. Rev. Lett. {\bf 83}
(1999) 3370; J. L. Hewett and M. Spriopulu, Ann. Rev. Nucl. Part.
Sci. {\bf 52}(2002)397.

\bibitem{lh1}
N. Arkani-Hamed, A. G. Cohen and H. Georgi, Phys. Lett. {\bf
B513}(2001)232; N. Arkani-Hamed, A. G. Cohen, T. Gregoire and J. G.
Wacker, JHEP {\bf 0208} (2002) 020, arXiv:hep-ph/0202089; N.
Arkani-Hamed, A. G. Cohen, E. Katz, A. E. Nelson, T. Gregoire and J.
G. Wacker, JHEP {\bf 0208}(2002) 021, arXiv:hep-ph/0206020; I. Low,
W. Skiba and D. Smith, Phys. Rev. {\bf D66}(2002)072001; D. E.
Kaplan and M. Schmaltz, JHEP {\bf 0310} (2003) 039,
arXiv:hep-ph/0302049.

\bibitem{lh2}
M. Schmaltz, Nucl. Phys. Proc. Suppl. {\bf 117}(2003)40; J. G.
Wacker, arXiv:hep-ph/0208235; S. Chang and J. G. Wacker,
arXiv:hep-ph/0303001; W. Skiba and J. Terning, Phys. Rev. {\bf D68}
(2003) 075001, arXiv:hep-ph/0305302.

\bibitem{lhest1}
T. Han, H. E. Logan, B. McElrath and L. T. Wang, Phys. Rev. {\bf
D67} (2003) 095004.

\bibitem{lhest2}
I. Low, W. Skiba and D.Smith, Phys. Rev. {\bf D66},(2002)072001, arXiv:hep-ph/0207243.

\bibitem{lhest3}
N. Arkani-Hamed, A. G. Cohen, E. Katz, A. E. Nelson, JHEP {\bf 0207}
(2002) 034, arXiv:hep-ph/0206021; S. Chang, JHEP {\bf 0312} (2003)
057, arXiv:hep-ph/0306034.

\bibitem{w3}
G. Burdman. M. Perelstein and A. Pierce, Phys. Rev. Lett. {\bf 90}
(2003) 241802; C. Dib, R. Rosenfeld and A. Zerwekh,
arXiv:hep-ph/0302068; T. Han, H. E. Logan, B. McElrath ans L. T.
Wang, Phys. Lett. {\bf B563} (2003) 191; Z. Sullivan,
arXiv:hep-ph/0306266.

\bibitem{smwork}
M. L. Ciccolini, S. Dittmaier and M. Kr\"amer, Phys. Rev. {\bf D68}
(2003) 073003, arXiv:hep-ph/0306234; B. A. Kniehl, Phys. Rev. {\bf D42}
(1990) 2253; B. A. Kniehl and C. P. Palisoc, Phys. Rev. {\bf D85} (2012)
75027.

\bibitem{width}
S.C. Park (KIAS), J. Song, Phys. Rev.{\bf D69},(2004) 115010,
arXiv:hep-ph/0306112v2.

\bibitem{fey}
T. Hahn,  Comput. Phys. Commun. {\bf 140} (2001) 418.

\bibitem{formcalc}
T. Hahn, M. Perez-Victoria, Comput. Phys. Commun. {\bf 118} (1999)
153.

\bibitem{pdfs}
J. Pumplin \textit{et al}., JHEP {\bf 0207}, (2002)012; D. Stump
\textit{et al}., JHEP {\bf 0310}, (2003) 046.

\bibitem{Stefan}
 R. K. Ellis and G. Zanderighi, JHEP {\bf 0802} (2008) 002.


\bibitem{OneTwoThree}
G.'t Hooft and M. Veltman, Nucl. Phys. {\bf B153} (1979) 365.

\bibitem{Four}
A. Denner, U Nierste and R Scharf, Nucl. Phys. {\bf B367} (1991)
637.

\bibitem{Five}
A. Denner and S. Dittmaier, Nucl. Phys. {\bf B658} (2003) 175.

\bibitem{TCPSS}
B. W. Harris and J.F. Owens, Phys. Rev. {\bf D65} (2002) 094032,
arXiv:hep-ph/0102128.

\bibitem{Lepage}
G. P. Lepage, J. Comput. Phys. {\bf 27} (1978) 192.


\bibitem{hdata}
K. Nakamura, \textit{et al}., J. of Phys. {\bf G37}, 075021 (2010).

\bibitem{range-1}
C. Csaki, J. Hubisz, G.D. Kribs, P. Meade, J. Terning, Phys. Rev.
{\bf D68} (2003) 035009; J. L. Hewett, F. J. Petriello,
T. G. Rizzo, JHEP {\bf 0310} (2003) 062; M. C. Chen, S. Dawson,
Phys. Rev. {\bf D70} (2004) 015003; M. C. Chen et al., Mod. Phys.
Lett. {\bf A21}(2006) 621; W. Kilian, J. Reuter, Phys. Rev. {\bf
D70} (2004) 015004.

\bibitem{CDF-ZH}
The ATLAS Collaboration, Phys. Lett. {\bf B700} (2011)163; The ATLAS
collaboration, Phys. Lett. {\bf B705} (2011)28; CMS Collaboration,
"Search for narrow resonances in dilepton mass spectra in pp
collisions at $\sqrt(s) = 7~TeV$", arXiv:1206.1849 [hep-ex].

\end{thebibliography}
\end{document}